\documentclass[
reprint,
 amsmath,amssymb,
 aps,
]{revtex4-2}

\usepackage{graphicx}
	\graphicspath{{fig/}}
\usepackage{dcolumn}
\usepackage{bm}

\usepackage{caption}
\usepackage{color}
\usepackage{hyperref}
\usepackage{comment}
\hypersetup{colorlinks=true, linkcolor=black, filecolor=black,   urlcolor=black,citecolor=black, anchorcolor=cyan}
\usepackage{algorithm2e}
	\SetKwComment{Comment}{/* }{ */}
	\RestyleAlgo{ruled}

\begin{document}

\preprint{APS/123-QED}

\title{Predator-prey survival pressure is sufficient to evolve swarming behaviors}
\thanks{Final version, to appear in New Journal of Physics, \\ DOI 10.1088/1367-2630/acf33a}%

\author{Jianan~Li}
 \email{lijianan@westlake.edu.cn}
 \affiliation{%
 	Department of Information Science \& Electronic Engineering, Zhejiang University, Hangzhou 310027, China \\ and School of Engineering, Westlake University, Hangzhou 310024, China
}%

\author{Liang~Li}%
 \email{lli@ab.mpg.de}
\affiliation{%
 Department of Collective Behavior, Max Planck Institute of Animal Behavior, 78464 Konstanz, Germany\\
 Center for the Advanced Study of Collective Behavior, Univ. of Konstanz, 78464 Konstanz, Germany\\
 and Department of Biology, University of Konstanz, 78464 Konstanz, Germany
}%

\author{Shiyu~Zhao}
 \email{zhaoshiyu@westlake.edu.cn}
\affiliation{
 Research Center for Industries of the Future, Westlake University, Hangzhou 310024,  China \\
 and School of Engineering, Westlake University, Hangzhou 310024,  China
}%


\begin{abstract}
The comprehension of how local interactions arise in global collective behavior is of utmost importance in both biological and physical research. Traditional agent-based models often rely on static rules that fail to capture the dynamic strategies of the biological world. Reinforcement learning has been proposed as a solution, but most previous methods adopt handcrafted reward functions that implicitly or explicitly encourage the emergence of swarming behaviors.
In this study, we propose a minimal predator-prey coevolution framework based on mixed cooperative-competitive multiagent reinforcement learning, and adopt a reward function that is solely based on the fundamental survival pressure, that is, prey receive a reward of $-1$ if caught by predators while predators receive a reward of $+1$.
Surprisingly, our analysis of this approach reveals an unexpectedly rich diversity of emergent behaviors for both prey and predators, including flocking and swirling behaviors for prey, as well as dispersion tactics, confusion, and marginal predation phenomena for predators. 
Overall, our study provides novel insights into the collective behavior of organisms and  highlights the potential applications in swarm robotics.
\end{abstract}

\maketitle


\section{Introduction} \label{sec:intro}

Swarming behaviors in the nature, such as starling flocks, fish schools, and sheep herds \cite{sumpter2010collective}, have been studied across various research fields, including biology \cite{krause2002living}, physics  \cite{vicsek1995novel}, and robotics \cite{chung2018survey}. 
Various agent-based models have been proposed to understand these collective behaviors. Most models are constructed based on designed static rules, including velocity alignment rules \cite{vicsek1995novel}, a balance of social forces including attraction and repulsion \cite{couzin2002collective,jia2019modelling}, and vision-based movement decisions \cite{pearce2014role, barberis2016large, lavergne2019group, bastien2020model}.
Although these models can phenomenally give rise to swarming behaviors including complex ring-shaped and swirling patterns \cite{chen2014minimal, jia2019modelling,liu2023modeling}, the interaction rules are mostly heuristic and static, failing to capture the adaptation property of the biological world.
%
To address this issue, recent studies have utilized evolution-based methods including neuroevolution \cite{olson2013predator} and reinforcement learning (RL) methods \cite{sunehag2019reinforcement, hahn2019emergent, durve2020learning, monter2023dynamics}, because they offer the potential for adaptable strategies, analogous to the way biological organisms evolve \cite{sutton2018reinforcement, muinos2021reinforcement, nasiri2022reinforcement}. 
In addition, multi-agent RL allows to model the interactions among individual agents in a swarm, and to optimize their collective behaviors \cite{sunehag2019reinforcement, hahn2019emergent, durve2020learning, monter2023dynamics}. 

However, the interaction mechanism or reward functions in evolution-based methods \cite{olson2013predator, sunehag2019reinforcement, hahn2019emergent, durve2020learning, monter2023dynamics} are task-specific \cite{kaiser2022innate, nelson2009fitness} meaning they are handcrafted by designers to intentionally fit for the characteristics implicitly or explicitly associated to swarming behaviors, which we refer to as \textit{swarm-dependent} in this study. 
For example, the reward function in \cite{durve2020learning} penalizes losing neighbors. That is, if an agent loses its neighbor, it receives a penalty of $-1$.
The recent work in \cite{monter2023dynamics} assumes prey receive larger reward as the area of domain of danger decreases \cite{hamilton1971geometry}, thus explicitly encouraging prey agents to move closer to each other.
The authors in \cite{olson2013predator} implicitly assume the attack efficiency is inversely proportional to the number of prey visible to the predator due to confusion. As a result, the number of prey in low-density areas reduces faster than that in high-density areas, resulting in a clustering phenomenon as anticipated. A similar confusion mechanism is used in \cite{hahn2019emergent}.

In this study, we propose a minimal predator-prey coevolution framework based on mixed cooperative-competitive multiagent reinforcement learning, where the reward function is solely based on the motivation to survive for both predators and prey. Specifically, prey receive a reward of $-1$ if caught by predators while predators receive a reward of $+1$. This reward function has no relation to objectives such as increasing neighbors, decreasing sparsity, enhancing alignment, or promoting other characteristics directly associated with swarming behaviors, thus \textit{swarm-independent}. 
Surprisingly, under the proposed framework, we find the simple survival pressure is sufficient to evolve flocking and swirling behaviors for prey. Quantitatively, we observe a steady increase in swarming density and group polarization. We also observe the emergent dispersion tactic, confusion and marginal predation phenomena for predators.

\section{Modeling}
\textbf{Environment}. 
We established a physics-based simulation environment where predators and prey interact with each other. The environment is a two-dimensional continuous space with two kinds of boundary conditions. 
As we will demonstrate in section \ref{sec:results}, distinct boundary conditions will encourage flocking or swirling behaviors.
The first kind is a finite square area, commonly used in previous studies such as \cite{mordatch2017emergence, lowe2017multi}. In this space, agents are unable to cross the boundaries, which are simulated as walls with a specified contact stiffness.
The second kind is periodic boundary condition where an agent passes through one side of the square environment re-appears on the opposite side with the same velocity. This treatment wraps around the edges of the environment into a torus, thus enabling us to approximate a large or infinite space. Periodic boundary condition is widely used in molecular dynamics simulation \cite{frenkel2001understanding} and swarm modeling \cite{vicsek1995novel, durve2020learning}. \\

\begin{figure}[t]
	\centering
	\includegraphics[]{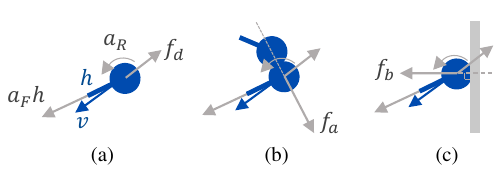}
	\caption{Illustration of active and passive forces.}
	\label{fig:force}
\end{figure}

\textbf{Agent Dynamics}. 
An agent, namely a predator or prey, is represented by a circle with a short line segment representing its heading, as shown in Fig.~\ref{fig:force}(a), where the unit vector $h \in \mathbb{R}^2, ||h||=1$ is the heading and $v \in \mathbb{R}^2$ is the velocity. The agents are subject to both active and passive forces. 

The active force is self-generated propulsion to drive ego-motion, and consists of two components as shown in Fig.~\ref{fig:force}(a).  
The first component is the action to move forward aligned with the heading and computed as $a_Fh  \in \mathbb{R}^2$ where $a_F \in \mathbb{R}$. The second component is the action to rotate its heading, and is denoted as $a_R \in \mathbb{R}$ within a threshold value.

The passive forces include dragging force $f_d\in \mathbb{R}^2$, elastic force between contact agents $f_a \in \mathbb{R}^2$ as shown in Fig.\ref{fig:force}(b), and elastic force between agents and boundaries $f_b \in \mathbb{R}^2$ as shown in Fig.\ref{fig:force}(c).
The dragging force is simply assumed to be in the opposite direction of velocity $v$ with its magnitude proportional to $||v||$. 
Elastic forces $f_a$ and $f_b$ follow Hooke's law, and sum up when an agent contacts multiple other agents or boundaries as $f_a=\sum_j f_{a,j}$ and $f_b=\sum_j f_{b,j}$. The elastic forces prevent agents overlap and reflect physical collision dynamics. 
It is remarked that the velocity may not be aligned with the heading direction when a collision happens.
In the simulation, the drag coefficient is set as $2$  N$\cdot$s/m and the contact stiffness coefficient is set as $50$ N/m.

Combining the aforementioned active and passive forces, the dynamics of an agent of species $i$ can be summarized as follows:
\begin{subequations} \label{eq:agentDynamics}
\begin{align}
	\dot x &= v ,\\
	\dot v &= (a_Fh + f_d + f_a + f_b)/m_i ,\\
	\dot \theta &= a_R,
\end{align}
\end{subequations}
where $x\in \mathbb{R}^2$ is the position, $m_i \in \mathbb{R^+}$ is the mass, $\theta \in (-\pi, \pi] \subset \mathbb{R}$ is the heading angle and $h=[\cos \theta, \sin\theta]^T$.
\\

\textbf{Agent Observation.} 
The observation model of an agent is assumed to be dependent on both metric and topological distance.  
Metric dependence means an agent can only perceive others in its perception range \cite{olson2013predator,sunehag2019reinforcement} which is assumed to be a disk with a pre-defined radius $R \in \mathbb{R^+}$ for simplicity.
Topological dependence refers to how many at most an agent can perceive concurrently rather than how far away. The threshold is set as $6$ which means an agent can perceive at most six allies and six adversaries. This setting is inspired by the work in \cite{ballerini2008interaction} that each bird interacts on average with six neighbors, and further confirmed in the recent work \cite{liu2023modeling}.
The information contained in an observation vector includes relative position and heading of observed agents for both allies and adversaries, and the agent's own position, velocity and heading. 
If the number of agents in the perception range exceeds the topological threshold, the farthest ones are removed, and if it does not reach the threshold, the rest part of the observation vector are masked out with zeros.
This topological modeling simplifies the structure of neural networks by fixing the input dimension.


\section{Framework of Coevolution}\label{sec:framework}
The optimization regime is set as a mixed cooperative-competitive multiagent reinforcement learning framework where predators and prey are learning and adapting their behaviors concurrently analogous to the coevolution in the nature. \\

\textbf{Homogeneity}. 
Agents of the same species in a swarm are assumed to be homogeneous.
This is reasonable when each agent has the same capability, responsibility, and goal as the others. More importantly, agents in a swarm typically display similar behavior patterns following the same interaction rule \cite{moussaid2011simple, pearce2014role, barberis2016large}. Homogeneity has been widely adopted in the modeling of swarm systems. For example, works such as those in \cite{vicsek1995novel,couzin2002collective, chen2014minimal, liu2023modeling, pearce2014role, barberis2016large, lavergne2019group, bastien2020model} propose singleton control laws that dominate all the agents in a swarm, while those in \cite{sunehag2019reinforcement, hahn2019emergent, durve2020learning, monter2023dynamics} employ parameter-sharing techniques for the agents' neural networks. As a result, the homogeneity leads to two features of the proposed training regime: parameter-sharing of actor-critic networks and replay buffer sharing among conspecifics, which are described as follows.\\

\textbf{Actor-Critic}.
Conspecifics share one critic and one actor network.
The critic network is used to evaluate the quality of an action taken by the agent, by estimating the expected future reward from that action, while the actor determines the agent's action $a=[a_F, a_R]^T$ based on its current observation. Together, the critic and actor work to improve the agent's behavior over time, by continuously refining its estimates of expected future rewards and adjusting its policy accordingly. 

The critic is designed to be decentralized allowing agents to evaluate based solely on local observations, without the knowledge of global states and actions like centralized critics used in \cite{lowe2017multi, huttenrauch2019deep}. This is analogous to the situation when living organisms can only perceive nearby surroundings with limited abilities.
Additionally, a decentralized critic provides better scalability as the number of agents in the system increases, making it particularly beneficial for swarm systems.

The actor is shared by conspecifics but not for their adversaries.  Policy-sharing technique has been widely used in traditional agent-based models \cite{vicsek1995novel,  couzin2002collective, chen2014minimal, liu2023modeling, pearce2014role, barberis2016large, lavergne2019group, bastien2020model}, and learning-based methods \cite{sunehag2019reinforcement, hahn2019emergent, durve2020learning, monter2023dynamics}. An advantage of policy-sharing is that the trained policy can be deployed across any number of agents of the same kind making it perfect for swarming research. 
It is remarked that although the parameters of the policy are shared, the output actions from the policy for each agent are different because they have different observations. \\

\textbf{Replay Buffer}. 
When an agent takes an action in a given state, it receives a reward and transitions to a next state. The replay buffer is used to store such transitions of the agent interacting with the environment in the form of tuples: state, action, reward and next state. 
The replay buffer serves to reduce correlations between consecutive experiences by exposing the agent to a more diverse set of experiences, thus helping to avoid over-fitting and stabilize the training process.

Two replay buffers $\mathcal{B}_0$ and $\mathcal{B}_1$ are used for predators and prey, respectively. There is no need to construct replay buffers for each individual because the homogeneity allows for interchangeable use of the experiences collected by conspecifics. Therefore, experiences collected from multiple conspecifics can be congregated into a single replay buffer, resulting in more resilient and effective learning outcomes. \\

\textbf{Rewards}.  
The reward for prey is simply set as $r = -1$ if it is caught by a predator, where the catch is represented by a contact between the two agents.
Compared to the swarm-dependent interaction mechanisms or reward functions in \cite{olson2013predator, sunehag2019reinforcement, hahn2019emergent, durve2020learning, monter2023dynamics}, the proposed reward function solely emphasizes the potential dangers of remaining in close proximity to predators, thus swarm-independent.
Similarly, the reward for a predator is $r = +1 $ if it catches prey. 
It is remarked that prey agents are not removed from the simulation after being caught by predators. The contact between predators and prey can be likened to a continuous process where predators extract energy from the prey while engaged in ``eating" them. Upon separation, the prey's survival reward returns to zero, signifying the cessation of energy transfer analogous to the termination of ``bleeding".
A decorative reward that mimics energy consumption due to movement is added to the essential survival reward, which is simply set as $-0.01|a_F| - 0.1 |a_R| $.
This reward function will cause the agent to exhibit laziness.

In the special case when boundaries exist in the environment, an additional penalty $-0.1$ is added to the reward function when contact between agents and boundaries happens.
This setting is designed to simulate either the presence of danger in the outside world or the situation where an agent chooses not to leave a specific location, such as a food-rich coral reef.
\\

\textbf{Algorithm}. 
The algorithm primarily adopts the multiagent deep deterministic policy gradient method as proposed in \cite{lowe2017multi, sutton1999policy}. However, certain modifications have been made to the original method, which are summarized as follows: 1) to account for the limited perception capabilities of agents, a decentralized critic has been employed in lieu of a centralized one; 2) agents of conspecifics share one actor network but not for their adversaries; 3) experiences collected by conspecifics are merged into a single replay buffer for more effective learning.

We designate the predators and prey as species $0$ and $1$, respectively. The dimensionality of the observation and action vectors are denoted as $d_o$ and $d_a$, respectively. We denote the discount factor, which determines the weight given to future rewards, as $\gamma$, and the soft update rate, which determines the speed at which a target network is updated towards a learning network, as $\tau$.
The proposed algorithm is summarized in Algorithm \ref{algo:ddpg}.

\begin{algorithm*}[t]
	\caption{Multiagent Deep Deterministic Policy Gradient Algorithm for Predator-Prey Coevolution}
	\tcp{$i=0$ for predators, $i=1$ for prey}
	
	\For {\textup{species $i$ = 0 to 1}}{
		Randomly initialize actor $\mu_i$ parametrized by $\theta^\mu_i$ and critic $Q_i$ parametrized by $\theta_i^Q$\;
		Initialize target actor $\mu'_i$  and target critic $Q'_i$, $\theta'^\mu_i \leftarrow \theta^\mu_i$,  $\theta'^Q_i \leftarrow \theta_i^Q$ \; 
	}
	\For{\textup{episode = 1 to M}}{
		Randomly spawn $n_0$ predators and $n_1$ prey; receive observations $o_i \in \mathbb{R}^{n_i\times d_o}$\;
		\For{\textup{t = 1 to max-episode-length}}{
			\textbf{for} \textit{all agents in species $i$}, 
			select actions $a_i = \mu_{\theta_i}(o_i)+ \mathcal{N}_t \in \mathbb{R}^{n_i\times d_a}$ where $\mathcal{N}_t$ is Gaussian noise;
			
			Execute actions $a_i$, receive reward $r_i\in \mathbb{R}^{n_i}$ and new observations $o'_i\in \mathbb{R}^{n_i\times d_o}$\;
			
			Store $(o_i, a_i, r_i, o'_i)$ 
			$\in$
			$( \mathbb{R}^{n_i \times d_o},
			\mathbb{R}^{n_i \times d_a},
			\mathbb{R}^{n_i},
			\mathbb{R}^{n_i \times d_o})$ 
			in replay buffer $\mathcal{B}_i$\;
			
			$o_i \leftarrow o'_i$\;
			
			Randomly sample a mini-batch of $S\in \mathbb{N}^+$ samples $(o_i^j, a_i^j, r_i^j, o'^j_i)$
			$\in$
			$( \mathbb{R}^{d_o},
			\mathbb{R}^{d_a},
			\mathbb{R},
			\mathbb{R}^{d_o})$ from $\mathcal{B}_i$\;
			
			Set $y^j_i=r^j_i + \gamma Q'_i(o_i'^j, a'^j_i)|_{a'^j_i=\mu'_i(o'^j_i)}$\;
			
			Update critics by minimizing the loss $\mathcal{L}(\theta^Q_i)= \frac{1}{S}\sum_{j=1}^S \left(y_i^j-Q_i(o^j_i, a^j_i)\right)^2 $\;
			
			Update actors using sampled policy gradient $\nabla_{\theta^\mu_i}J \approx 
			\frac{1}{S} \sum_{j=1}^S \nabla_{\theta^\mu_i}\mu_i(o_i^j)
			\nabla_{\mu_i(o_i^j)} Q_i (o_i^j, \mu_i(o_i^j))$\;
			
			Soft-update target network parameters 
			$\theta'^\mu_i \leftarrow \tau\theta^\mu_i + (1-\tau)\theta'^\mu_i$, 
			$\theta'^Q_i \leftarrow \tau\theta^Q_i + (1-\tau)\theta'^Q_i$\;
		}	
	}
	\label{algo:ddpg}
\end{algorithm*}

\section{Results and Analysis}\label{sec:results}

\textbf{Simulation Setup}. In the training phase, we instantiate $n_0=3$ predators and $n_1=10$ prey in the environment.
This selection of population size is based on the assessment of hardware performance, as having more agents in the environment requires longer computation time.
Nevertheless, the population size can be arbitrary as the policy is decentralized. 
In the evaluation phase, we deploy the trained policy on $50$ prey, to yield a more distinct visual effect of swarming behaviors, whilst maintaining the same number of predators.

\begin{figure}[t]
	\centering
	\begin{minipage}{0.43\linewidth}
		\centering
		\fbox{\includegraphics[width=1\linewidth]{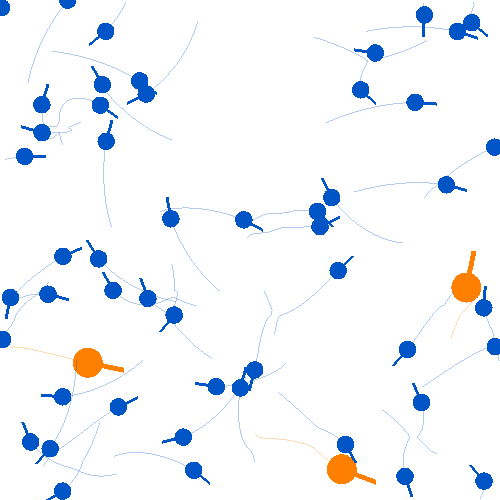}}
	\end{minipage}
	\hspace{1em}
	\begin{minipage}{0.43\linewidth}
		\centering
		\fbox{\includegraphics[width=1\linewidth]{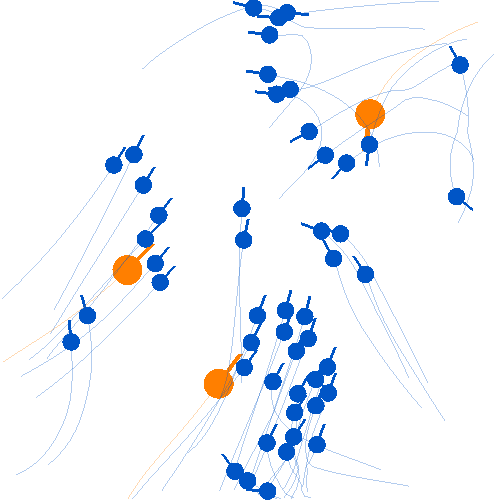}}
	\end{minipage}
	\caption{(a) Before coevolution. (b) After coevolution.}
	\label{fig:evolution}
	\vspace{1em}
	\includegraphics[]{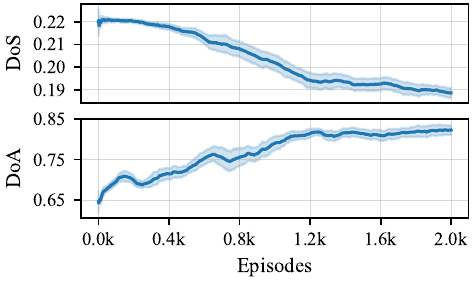}
	\caption{Episodic evolution of DoS and DoA.}
	\label{fig:standard}
	\vspace{1em}
	\includegraphics[]{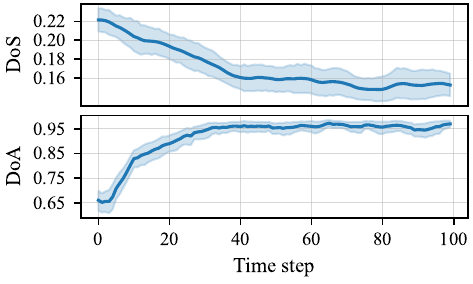}
	\caption{Time evolution of DoS and DoA in an episode by loading trained policies on agents.}
	\label{fig:standardAfterTrain}
\end{figure}

\begin{figure}[t]
	\centering
	\begin{minipage}{0.43\linewidth}
		\fbox{\includegraphics[width=1\linewidth]{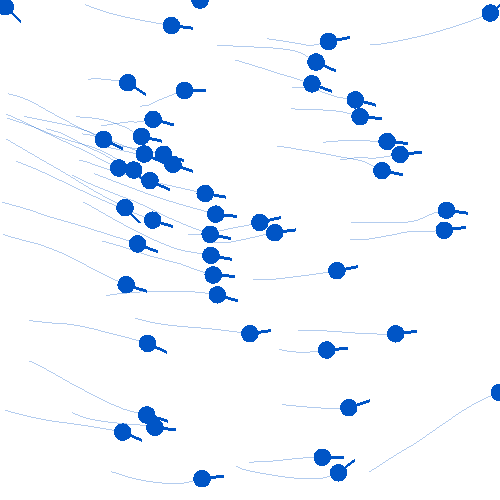}}
	\end{minipage}
	\hspace{1em}
	\begin{minipage}{0.43\linewidth}
		\fbox{\includegraphics[width=1\linewidth]{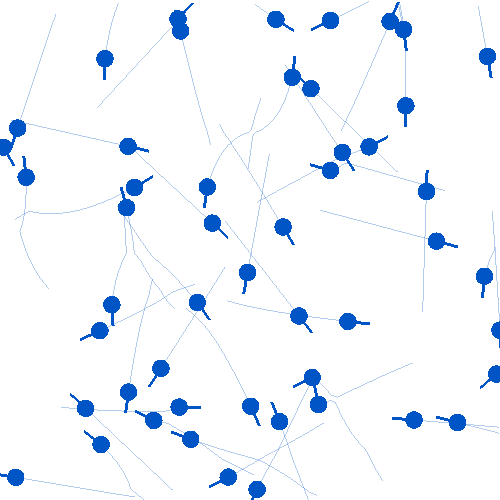}}
	\end{minipage} 
	\caption{(a) Remove predators after training with predators. (b) Remove predators in training.}
	\label{fig:withoutPred}
	\vspace{1em}
	\begin{minipage}{0.43\linewidth}
		\fbox{\includegraphics[width=1\linewidth]{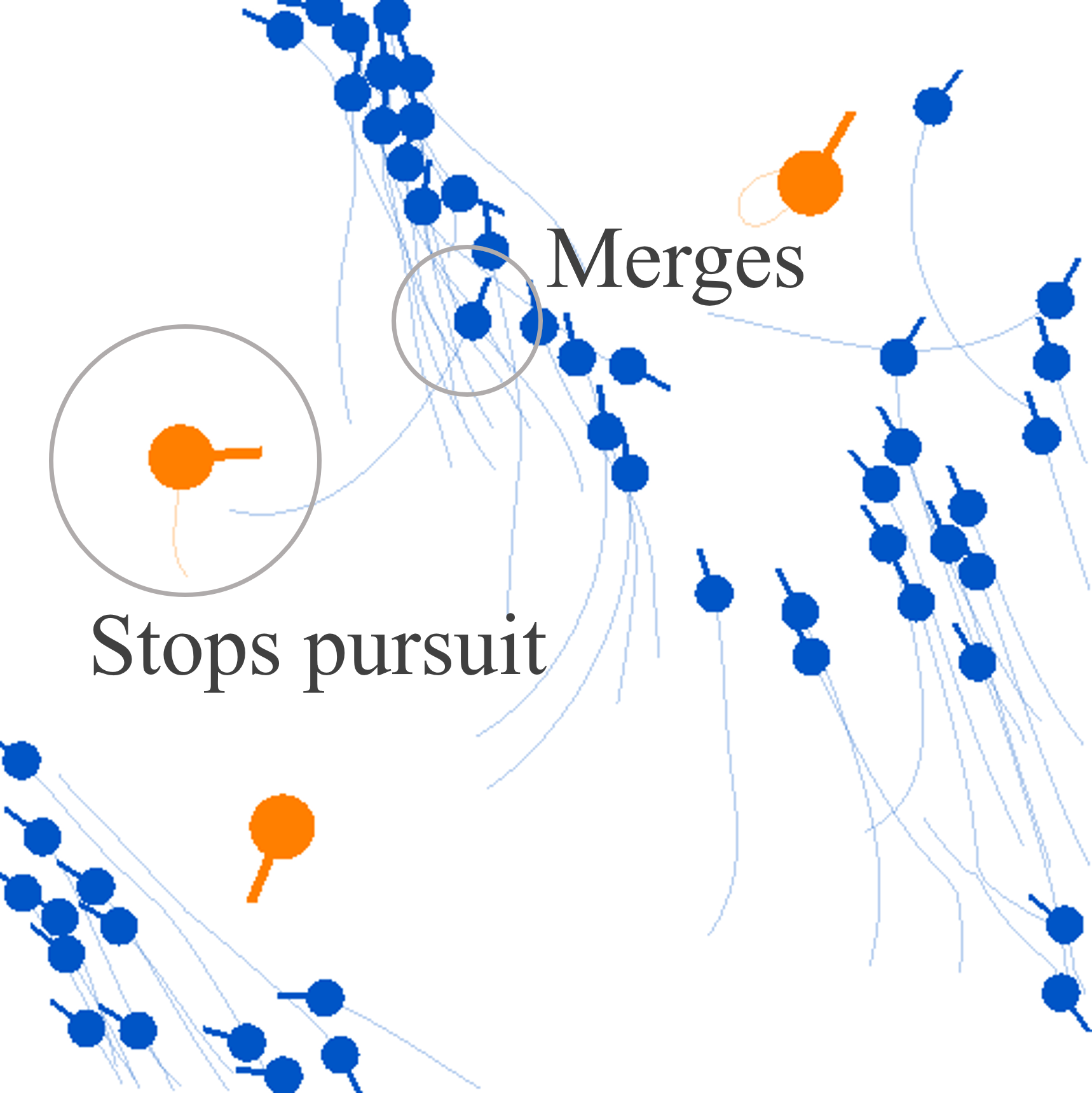}}
	\end{minipage}
	\hspace{1em}
	\begin{minipage}{0.43\linewidth}
		\fbox{\includegraphics[width=1\linewidth]{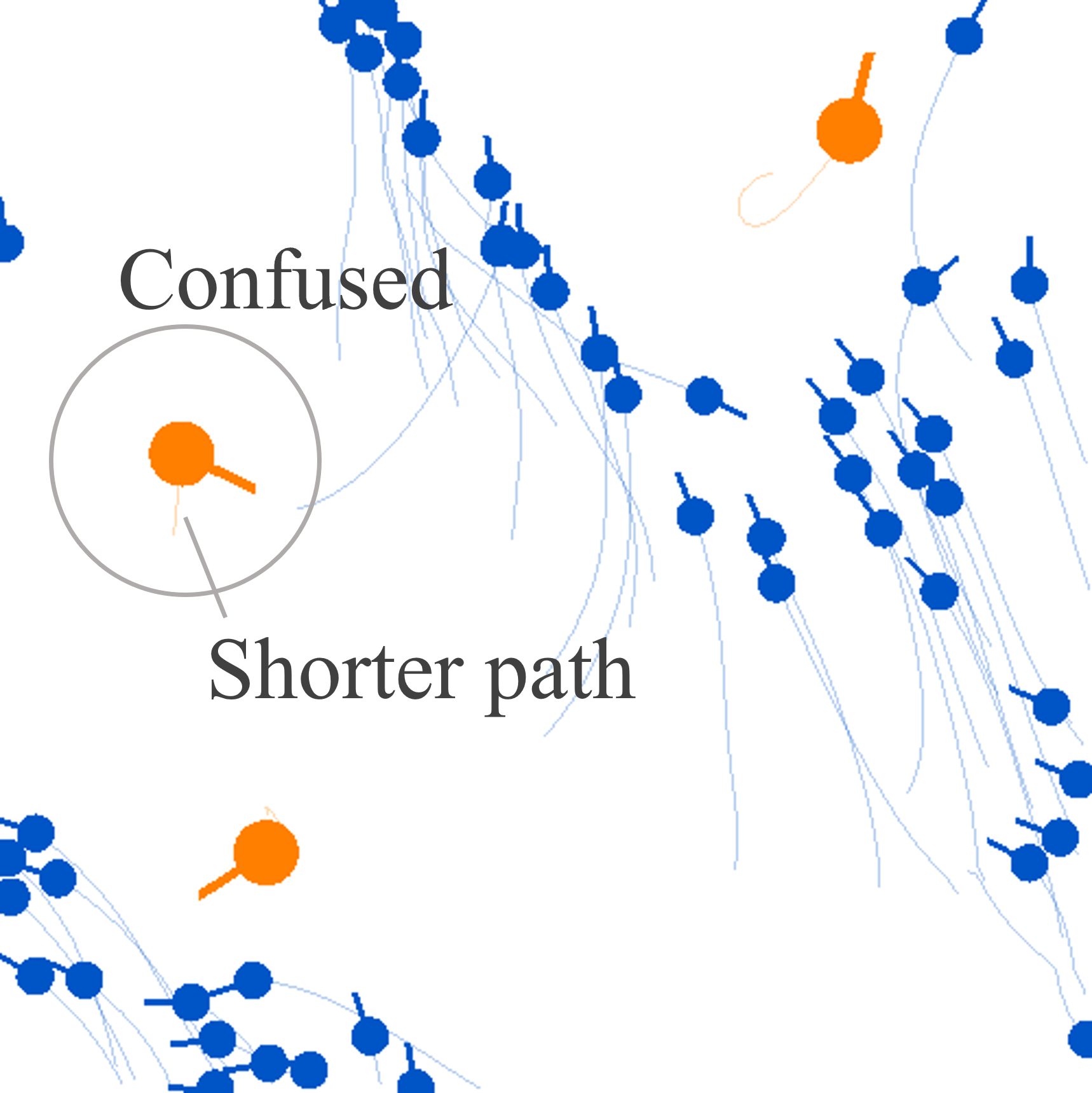}}
	\end{minipage} 
	\caption{Confusion effect. (a) t=30. (b) t=33.}
	\label{fig:confuse}
	\vspace{1em}
	\begin{minipage}{0.43\linewidth}
		\fbox{\includegraphics[width=1\linewidth]{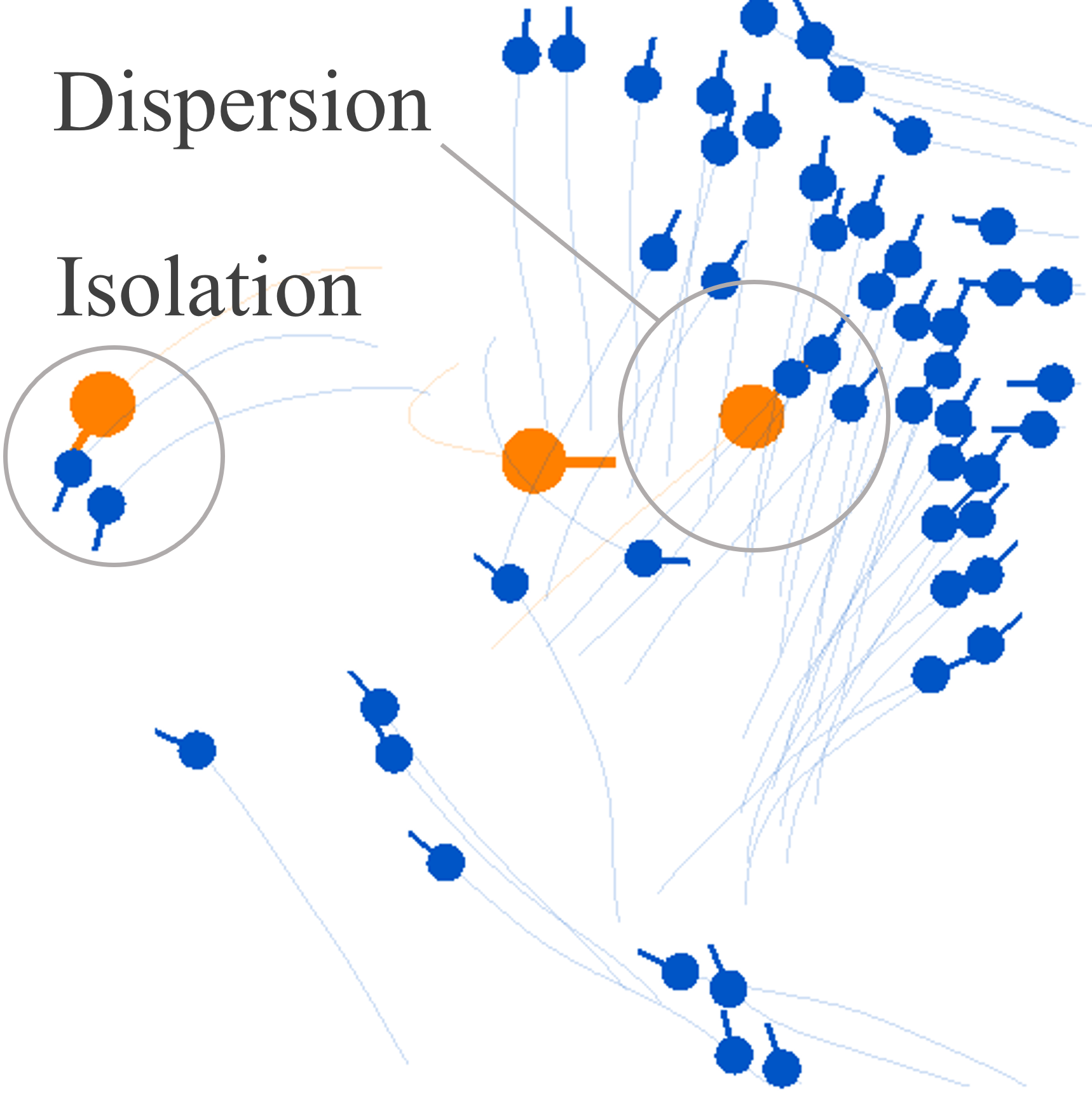}}
	\end{minipage}
	\hspace{1em}
	\begin{minipage}{0.43\linewidth}
		\fbox{\includegraphics[width=1\linewidth]{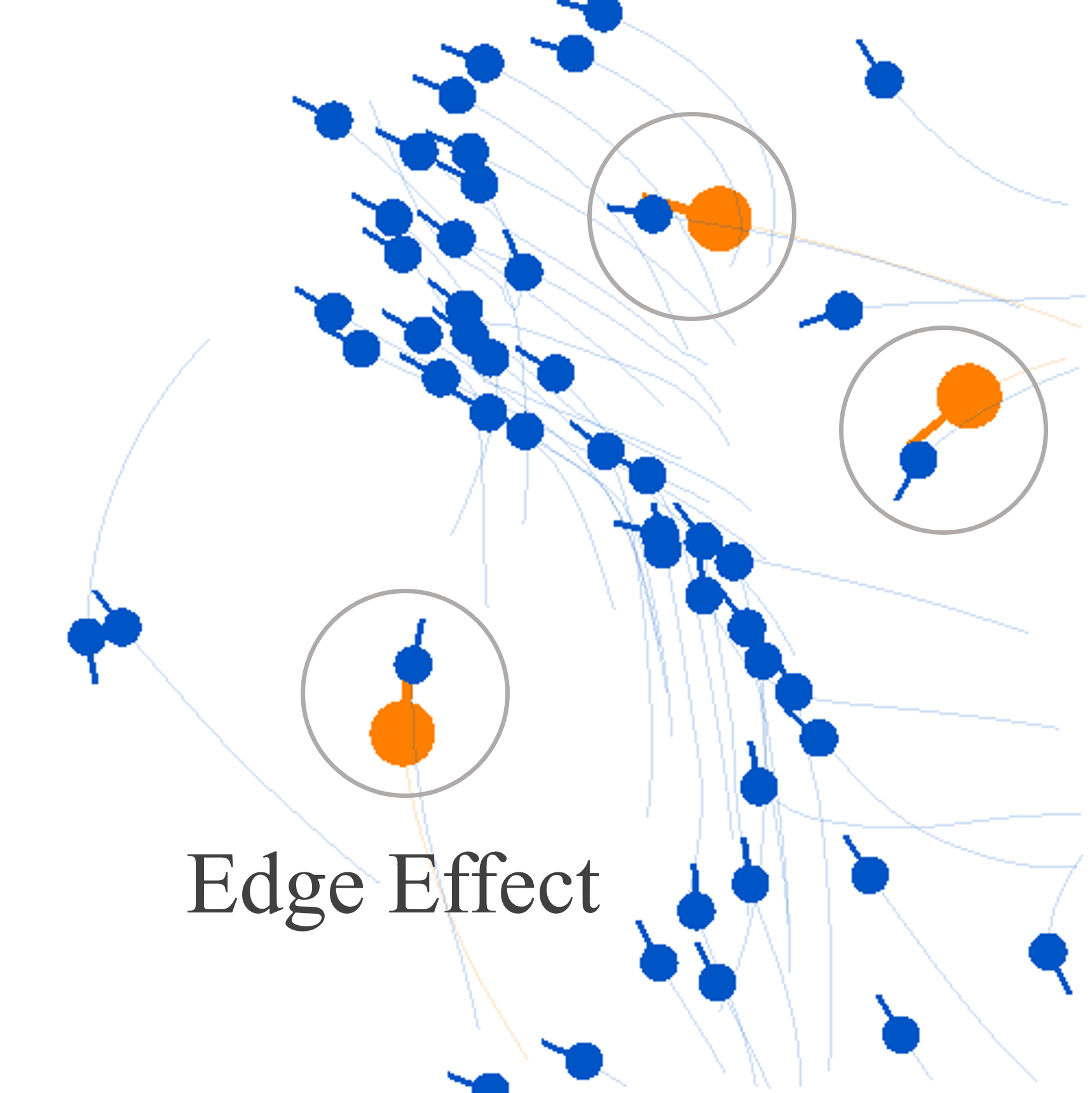}}
	\end{minipage}
	\caption{(a) Dispersion tactic. (b) Edge effect.}
	\label{fig:edgeEffect}
	\vspace{1em}
	\begin{minipage}{0.43\linewidth}
		\fbox{\includegraphics[width=1\linewidth]{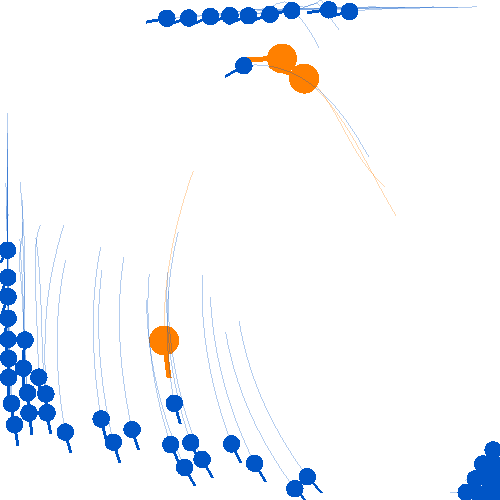}}
	\end{minipage}
	\hspace{1em}
	\begin{minipage}{0.43\linewidth}
		\fbox{\includegraphics[width=1\linewidth]{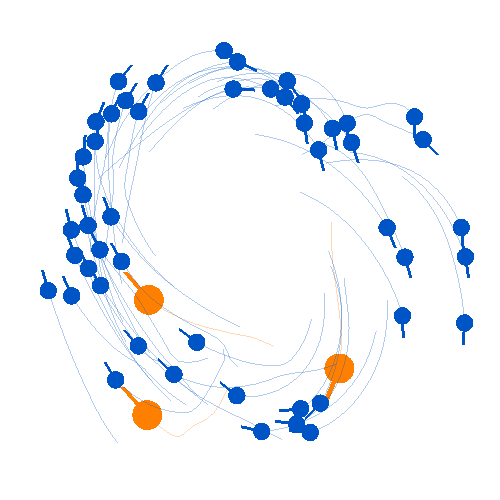}}
	\end{minipage}
	\caption{In a confined space. (a) Without boundary penalty. (b) With boundary penalty.}
	\label{fig:confined}
\end{figure}

In order to provide a quantitative assessment of swarming behaviors, we introduce two measures: degree of sparsity (DoS) and degree of alignment (DoA).
The $\text{DoS} \in [0,1] \subset \mathbb{R} $ is defined as the average normalized distance to the nearest neighborhood of all conspecifics in an episode as
\begin{equation}\label{eq:DoS}
	\text{DoS} = \frac{1}{TND} \sum_{t=1}^{T}\sum_{j=1}^{N}|| x_j(t) - x_k(t)||
\end{equation}
where $x_j(t)$ is $j$-th agent position at time step $t$, $k=\arg\min ||x_j(t) - x_k(t)||, {k\in \{1,2,...,N\}\backslash j}$, $T\in\mathbb{N}^+$ is episode length, $N\in\mathbb{N}^+$ is the total number which is equal to $n_1$ for prey, $D  \in \mathbb{R}^+$ is the environment size defined as the maximum possible distance for two agents. For example, for a periodic square environment with edge length $2$, the largest possible distance is $\sqrt 2$. 
The symbol $||\cdot||:  \mathbb{R}^2 \rightarrow  \mathbb{R}$ denotes the Euclidean norm mapping.
A smaller DoS indicates a denser swarm. An extreme case is when all conspecifics aggregate at the same point resulting in zero DoS.
The definition of $\text{DoA} \in [0,1] \subset \mathbb{R}$ is
\begin{equation}\label{eq:DoA}
	\text{DoA} = \frac{1}{2TN} \sum_{t=1}^{T}\sum_{j=1}^{N} ||h_j(t) + h_k(t)||
\end{equation}
where $h_j \in \mathbb{R}^2$ is the heading of $j$-th agent, $k$ is the same as in the definition of DoS.
It is worthy to remark that DoA is not equivalent to the mean heading of all conspecifics.
This is because although the headings are similar in the same flock resulting in a high DoA, multiple flocks with different headings may cancel each other, as shown in Fig.~\ref{fig:evolution}(b) and Fig.~\ref{fig:confined}(b). 
Therefore, calculating the relative quantity within a local neighborhood is more appropriate. A similar conclusion can be drawn for DoS. \\

\textbf{Emergent Flocking Behavior}.
A typical scenario before and after evolution are shown in Fig.~\ref{fig:evolution}(a) and Fig.~\ref{fig:evolution}(b), respectively. The predators are depicted in orange, while the prey are in blue and a smaller size. 
Prior to evolution, the prey move randomly in different directions due to randomly initialized policy. Similarly, predators exhibit purposeless movements without the intention to pursue prey.
%
The agents' behavior has undergone substantial changes after $2000$ episodes of coevolution, as depicted in Fig.~\ref{fig:evolution}(b). 
Notably, the prey exhibit a remarkable emergence of cohesive movement patterns and a high degree of alignment within multiple flocks, adhering to the well-known flocking basic rules outlined in \cite{reynolds1987flocks}.

The episodic evolution of running average of DoS and DoA is shown in Fig.~\ref{fig:standard}, where the running average length is $100$-episode and the shaded area indicates $95\%$ confidence interval.
Specifically, the DoS drops steadily from $22\%$ of the environment size to around $19\%$ suggesting a more cohesive movement of prey. Meanwhile, the DoA increases from $0.65$ to approximately $0.82$ indicating any two prey in the vicinity exhibit a higher degree of alignment.
It is remarked that the initial value of DoA is about $0.65$. This is because for a uniformly distributed heading angle ranging from $-\pi$ to $\pi$, the expected DoA is given by $\mathbb{E}[\cos (\phi/2)] = 2/\pi \approx 0.64$, where $\phi$ denotes the angle formed by two headings. This expected value is quite close to the value read from Fig.~\ref{fig:standard}.

The DoS and DoA shown in Fig.~\ref{fig:standard} is a mean value computed from the entire episode, as indicated by $1/T\sum_{t=1}^{T}(\cdot)$ in Eq.~\ref{eq:DoS} and \ref{eq:DoA}.
For established flocks, as shown in Fig.~\ref{fig:evolution}(b) or Fig.~\ref{fig:edgeEffect}, the DoS is lower than $19\%$ while its DoA exceeds $0.82$.
This can be shown by loading trained policies into agents, and plotting the DoS and DoA as functions of \textit{time steps} in an episode, as shown in Fig.~\ref{fig:standardAfterTrain}.
In this case, the DoS decreases to approximately $15\%$ while the DoA reaches around $0.96$. 

It is possible to question whether the observed alignment and distance reduction between neighboring prey is a result of swarming or merely a consequence of prey fleeing from predators in the same direction, which is commonly referred to as \textit{herding}. To differentiate between these two phenomena, we conducted additional simulations where only trained prey agents are present. A typical scenario is illustrated in Fig.~\ref{fig:withoutPred}(a), where we observe that the prey still exhibit a high degree of alignment even in the absence of predators, which is comparable to the well-known Vicsek model \cite{vicsek1995novel}. This finding further supports the hypothesis that swarming behavior observed in the absence of predators is a strategy employed by prey to avoid predators.

These findings are noteworthy because they reveal how simple predator-prey interaction, driven solely by the motivation to survive, could give rise to conspicuous flocking behavior that exhibits both cohesion and alignment characteristics.
This leads us to hypothesize that the emergent flocking behavior is largely an outcome of passive space extrusion and polarization induced by predators.
\\

\textbf{Emergent Confusion Effect, Dispersion Tactic and Edge Effect}.
During the pursuit, predators exhibit confusion in certain situations, as illustrated in Fig.~\ref{fig:confuse}(a)(b) at time step $30$ and $33$ of an episode. When the prey agent merges into a flock, the predator gives up the chase, slows down, and stagnates for a while as evidenced by a shorter path, appearing confused and uncertain about which prey to focus on.

Additionally, as shown in Fig.~\ref{fig:edgeEffect}(a), predators often employ a \textit{dispersion tactic}, in which they first move towards the center of a swarm to disperse it and then focus on isolated ones to capture \cite{huffaker1958experimental, major1978predator}.
The resulting isolation or the phenomenon that predators frequently catch prey on the periphery of a swarm is referred to as \textit{marginal predation} or \textit{edge effect} \cite{duffield2017marginal,romenskyy2020quantifying}, as shown in Fig.~\ref{fig:edgeEffect}(b). 
The observed marginal predation behavior suggests that predators may be less confused when targeting prey on the periphery of a swarm, potentially due to a smaller number of prey resulting in an elevated predation rate \cite{major1978predator, krakauer1995groups}, or a higher encounter rate \cite{duffield2017marginal}. 
Overall, the emergent confusion effect and marginal predation behavior highlight the challenges predators face when trying to select the optimal target from a group of prey. \\

\textbf{Emergent Swirling Behavior}.
We now examine the scenario when the environment is confined, and consider two cases: without and with penalties when prey collide with boundaries.
It is remarked that the extra boundary penalty will complicate the environment, thus slowing down the learning process. 
To accelerate the evolution of prey, we endow the predators a behavioral rule that creates survival pressure by directly moving towards their nearest prey. This rule controls two active forces: first, the predator rotates its heading to point towards the nearest prey, and then it moves directly towards the target at maximum speed. By following this rule, the predators exert survival pressure on the prey and drive their evolutionary adaptation.

Without penalty, there are no significant differences in flocking behaviors compared to the infinite space case, except for boundary aggregation, as in Fig.~\ref{fig:confined}(a), similar to fish swimming in a fishbowl or pond.
With penalty, an enigmatic swirling behavior emerges as in Fig.~\ref{fig:confined}(b). Such circular motion has been observed in fish and insects, yet it is hardly understood and still remains unclear in the scientific community \cite{parrish1999complexity, franks2016social}.
We hypothesize that the extra penalty acts as a disincentive for prey to leave a food-rich location, such as coral reefs, or a response to perceived threats in the outside world. Swirling may be an optimal tactic for prey to remain at the same place while simultaneously evading potential predators.\\

\textbf{Effect of Speed Limit Ratio}.
In prior studies, we assume a maximum speed ratio $||v_0||_{\max}:||v_1||_{\max} = 1:1$ between predators and prey.
This assumption is reasonable when predator and prey species have similar maximum speeds. Here, we investigate the emergent phenomenon when their maximum speeds have a distinct difference.

\begin{figure}[t]
	\centering
	\includegraphics[]{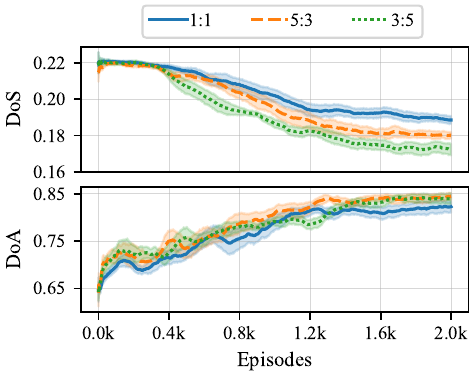}
	\caption{Effect of speed limit ratio.}
	\label{fig:effectSpeed}
\end{figure}

As shown in Fig.~\ref{fig:effectSpeed}, adjusting the speed limit ratio to $5:3$, we observe more pronounced swarming characteristics evidenced by a smaller DoS dropping from $19\%$ to $18\%$ and a larger DoA rising from $0.82$ to around $0.85$.
Further, tuning the speed limit ratio to $3:5$ results in a higher convergence rate of DoS and a lower value of around $17\%$, attributed to prey's athletic ability to evade predators while still maintaining formation. \\

\begin{figure}[t]
	\includegraphics[]{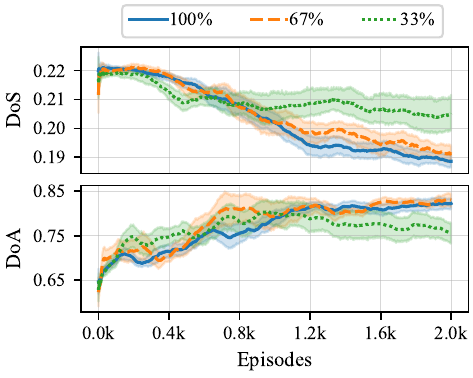}
	\caption{Effect of perception range.}
	\label{fig:effectFoV}
\end{figure}

\textbf{Effect of Perception Range}.
In other simulations, predators and prey are assumed to have a perception range equal to the environment size, that is, $R=D$, which is reasonable when organisms have a perception range much larger than their body length. 
Here, we intentionally tune the perception range to be $R=2/3D$ and $R=1/3D$ to investigate its effect on flocking behaviors, as shown in Fig.~\ref{fig:effectFoV}.

Smaller perception ranges lead to less pronounced flocking behavior, as indicated by a larger DoS and smaller DoA. This effect is especially notable when the perception range is only one-third of the environment size, where DoS increases from $19\%$ to around $20.5\%$, and DoA drops from $0.82$ to $0.75$. 
These findings suggest that perception range plays a crucial role in facilitating the emergence of flocking behaviors. \\

\begin{figure}[t]
	\includegraphics[]{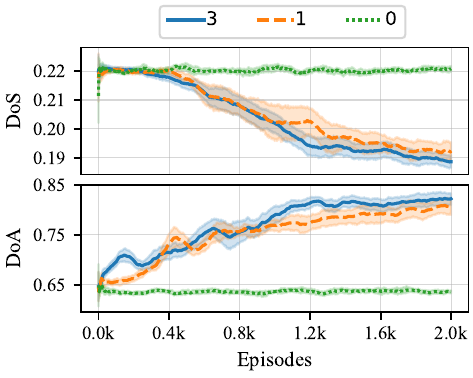}
	\caption{Effect of number of predators.}
	\label{fig:effectNumPred}
\end{figure}

\textbf{Effect of Number of Predators}.
In previous simulations, the number of predators present in training is set as $n_0=3$. Here, we specifically investigate the special cases where the number of predators during evolution are $n_0=0$ and $n_0=1$. The episodic evolutions of DoS and DoA are shown in Fig.~\ref{fig:effectNumPred}.

Comparing the cases of a single predator and three predators, we observe that the emergence of swarming behaviors is slightly slower with a single predator. This can be explained by the greater survival pressure exerted by multiple predators, which accelerates the prey's evolution.
In the special case \textit{without} predators in the evolution, it can be seen that the DoS and DoA of the prey remain unchanged, indicating that no swarming phenomenon emerges, as exemplified in Fig.~\ref{fig:withoutPred}(b). This evidence, from another perspective, further corroborates our hypothesis that predator-prey survival pressure is sufficient to promote swarming behaviors. 


\section{Discussion and Conclusion}
In this article, a minimal predator-prey coevolution framework based on mixed cooperative-competitive multiagent RL is proposed, with a swarm-independent reward function based solely on the motivation to survive. We have observed the emergent flocking and swirling behaviors for prey, and dispersion tactic, confusion and marginal predation phenomena for predators.
Based on these findings, we hypothesize swarming behaviors can be largely an outcome of passive space extrusion and polarization induced by predators. At the same time, predators may face challenges when trying to select their optimal target from a group of prey.
While the proposed framework may not perfectly match the real evolution mechanism in the nature and may not depict the evolution process of all kinds of swarming organisms, it provides a feasible approach for swarming research without the need for handcrafted reward function design. 

With the design philosophy of \textit{minimalism}, the proposed framework could serve as a starting point for further experimentation and customization.
For instance, one could perform a quantitative analysis on the effect of observation noise, or introduce a third-party species to study emergent behaviors.
In the field of robotics, collision penalties can be further incorporated into reward functions to realize collision avoidance simultaneously, enabling robots to exhibit swarming behaviors in a more natural manner without complex handcrafted interaction rules.
Overall, the proposed framework and findings could contribute to a better understanding on swarm intelligence of organisms and physical active matter, and have potential applications in swarm robotics.

\appendix
\section{Video Abstract and Animations}
The video abstract and simulation animations of Fig.\ref{fig:evolution}, \ref{fig:withoutPred}, \ref{fig:confuse}, \ref{fig:edgeEffect} and \ref{fig:confined} can be found in the supplementary materials.

\section{Implementation Details}
We provide more implementation details about the established predator-prey environment and the proposed coevolution algorithm.
In the numerical simulation, the agent dynamics described in Eq.~\ref{eq:agentDynamics} are performed at discrete time steps. For an agent of species $i$, its states at time step $t+1$ are updated in the following order:
\begin{align*}
	\theta(t+1) &= \theta(t) + a_R\Delta t, \\
	v(t+1) &= v(t) + \left(a_Fh + f_d + f_a + f_b\right)\Delta t/m_i, \\
	x(t+1) &= x(t) + v(t) \Delta t,
\end{align*}
where $\Delta t$ is the time step with its value shown in Table~\ref{tab:env_parameters}.
In the periodic boundary condition, when an agent moves beyond one edge of the environment, its position is further updated such that it reappears on the opposite edge as if the edges were connected.  For example, if the agent has moved beyond the maximum $x$-coordinate, we then set its $x$-coordinate to the minimum $x$-coordinate plus the distance it has moved beyond the maximum $x$-coordinate.
The agent's actions $a_F$ and $a_R$ are the outputs of the actor neural network, and re-scaled to fit within specified ranges as shown in Table~\ref{tab:env_parameters}. Specifically, $a_F$ ranges from zero to its maximum linear acceleration, while $a_R$ ranges from negative maximum angular velocity to positive maximum angular velocity.
To calculate $f_a$, we first analyze whether any two agents collide based on their sizes and distances. If collision occurs, we then apply Hooke's law to calculate the elastic forces resulting from the deformation, and sum them up as $f_a = \sum_j f_{a,j}$ if multiple collisions happen. The procedure of calculating $f_b$ is the same.
\begin{table}[ht] 		
	\caption{Environment parameters}
	\label{tab:env_parameters}
	\centering
	\begin{tabular}{lll}
		\hline\noalign{\smallskip}
		Parameter  & Value & Unit \\
		\noalign{\smallskip}\hline\noalign{\smallskip}
		Mass of predator  & 1 & kg 	\\
		Mass of prey   & 1 & kg 	\\
		Max speed     & 0.5 or 0.3 & m/s \\
		Max linear acc.  & 1 & m/s$^2$ \\
		Max angular vel. & 0.5 & rad/s \\
		Env edge length & 2 & m \\
		Contact stiffness  & 50 & N/m           \\
		Drag coefficient  & 2 & N$\cdot$s/m \\
		Time step  & 0.1 & s \\
		\noalign{\smallskip}\hline
	\end{tabular}
\end{table}

The observation vector for an agent has the following form:
\begin{equation*}
\begin{bmatrix}
	\text{agent's own pos., vel. and heading,} \\ 
	\text{relative pos. and headings of observed predators,} \\
	\text{relative pos. and headings of observed prey}
\end{bmatrix}
\end{equation*}
The relative positions and headings are reordered from the nearest to the farthest based on range. This is reasonable as conspecifics are considered as homogeneous, as explained in section \ref{sec:framework}. 
If the number of observed agents exceeds the topological limit, the farthest ones are removed, and if it does not reach the limit, the rest part of the observation vector is masked out with zeros.

\begin{figure}[t]
	\centering
	\includegraphics[]{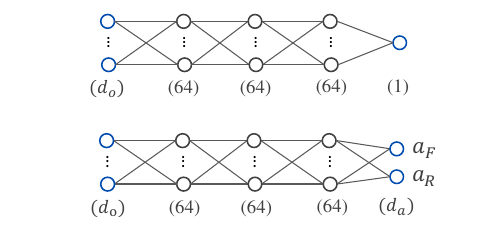}
	\caption{Illustration of critic and actor neural networks.}
	\label{fig:networks}
\end{figure}

Both critic and actor are encoded by deep feed-forward neural networks with rectified linear unit (ReLU) activation with an input dimension $d_o$ equivalent to the length of the observation vector. Each network consists of three hidden layers with $64$ neurons per layer as shown in Fig.~\ref{fig:networks}, where $d_a=2$ is the output dimension of the actor network and $a_F,a_R$ are the output actions.

For a detailed introduction to RL and multi-agent RL, we refer to \cite{sutton2018reinforcement, zhao2024reinforcement} and \cite{zhang2021multi}, respectively. For an introduction to multiagent deep deterministic policy gradient algorithm, we refer to \cite{lowe2017multi}. 
In the training, we first randomly initialize the actor network $\mu_i$ parameterized by $\theta_i^\mu$ and the critic network $Q_i$ parameterized by $\theta_i^Q$ for species $i$. 
In this study, predators and prey are denoted as species $i=0$ and $1$, respectively. 
The corresponding target actor $\mu'_i$ and critic  $Q'_i$ are also initialized to help mitigate the issue of non-stationarity by slowly updating the value functions as policy improves. 
At the beginning of each episode, $n_0 \in \mathbb{N^+}$ predators and $n_1 \in \mathbb{N^+}$ prey are placed in the environment at random positions with random headings.
Based on current observations $o_i \in \mathbb{R}^{n_i\times d_o}$ where $d_o$ is the dimension of the observation vector, the agents perform actions $a_i \in \mathbb{R}^{n_i\times d_a}$ according to current actors where $d_a$ is the dimension of the action, receive rewards $r_i \in \mathbb{R}^{n_i}$, and obtain new observations $o'$. 
The resulting experience tuple $(o_i,a_i,r_i,o'_i)$ are saved into replay buffer $\mathcal{B}_i$. 

In learning, agents draw a random mini-batch of $S\in \mathbb{N}^+$ samples from $\mathcal{B}_i$, and each sample is denoted as 
$(o_i^j, a_i^j, r_i^j, o'^j_i)$
$\in$
$( \mathbb{R}^{d_o},
\mathbb{R}^{d_a},
\mathbb{R},
\mathbb{R}^{d_o})$. 
The critic is updated by minimizing the loss:
\begin{equation*}
	\mathcal{L}(\theta^Q_i)= \frac{1}{S}\sum_{j=1}^S \left(y_i^j-Q_i(o^j_i, a^j_i)\right)^2 
\end{equation*}
where the target value $y^j_i=r^j_i + \gamma Q'_i(o_i'^j, a'^j_i)$, $a'^j_i=\mu'_i(o'^j_i)$.
To update the actor, the policy gradient theorem \cite{sutton1999policy} is used, and the gradient is approximated as:
\begin{equation*}
	\nabla_{\theta^\mu_i}J \approx 
	\frac{1}{S} \sum_{j=1}^S \nabla_{\theta^\mu_i}\mu_i(o_i^j)
	\nabla_{\mu_i(o_i^j)} Q_i (o_i^j, \mu_i(o_i^j))
\end{equation*} 
Finally, the target networks are soft-updated accordingly.
The exploration rate $\epsilon$ is the probability that the agent will choose to explore the environment instead of exploiting it. In this study, $\epsilon$ is set to gradually decrease with each episode achieved by using the formula $\max(0.05,~ \epsilon - \text{5e-5})$. 
Similarly, the action noise $\mathcal{N}_t$ is also set to decrease gradually with each episode as $\max(0.05, ~\mathcal{N}_t - \text{5e-5})$, with their initial values shown in Table~\ref{tab:algo_parameters}.
The learning process is carried out until a state of dynamic equilibrium is achieved between the predators and prey, such that neither party can obtain their future rewards by altering their respective policies.
The hyper-parameters for the proposed algorithm are summarized in Table~\ref{tab:algo_parameters}.
\begin{table}[ht] 		
	\caption{Hyper-parameters of algorithm}
	\label{tab:algo_parameters}
	\centering
	\begin{tabular}{ll}
		\hline\noalign{\smallskip}
		Hyper-parameter & Value \\
		\noalign{\smallskip}\hline\noalign{\smallskip}
		Number of episodes         & 2000 \\
		Episode length          & 100    \\
		Number of hidden layers    & 3    \\
		Hidden layer size		& 64   \\
		Learning rate of actor   & 1e-4 \\
		Learning rate of critic  & 1e-3 \\
		Discount factor          & 0.95 \\
		Soft-update rate         & 0.01 \\
		Initial exploration rate & 0.1 \\
		Initial noise rate & 0.1 \\
		Replay buffer size & 5e5 \\
		Batch size  & 256 \\
		\noalign{\smallskip}\hline
	\end{tabular}
\end{table}

\section*{Acknowledgments}
This work was supported in part by the STI 2030-Major Projects (Grant No. 2022ZD0208804) and the Hangzhou Key Technology Research and Development Program (Grant No. 20200416A16).
L.L. acknowledges funding support from the Max Planck Society, the Deutsche Forschungsgemeinschaft (DFG, German Research Foundation) under Germany’s Excellence Strategy-EXC 2117–422037984, the Sino-German Centre in Beijing for generous funding of the Sino-German mobility grant M-0541, and Messmer Foundation Research Award.
The authors would also like to thank the anonymous reviewers for their valuable insights and feedback.

\bibliography{Bibliography}

\providecommand{\noopsort}[1]{}\providecommand{\singleletter}[1]{#1}%
\begin{thebibliography}{40}%
\makeatletter
\providecommand \@ifxundefined [1]{%
 \@ifx{#1\undefined}
}%
\providecommand \@ifnum [1]{%
 \ifnum #1\expandafter \@firstoftwo
 \else \expandafter \@secondoftwo
 \fi
}%
\providecommand \@ifx [1]{%
 \ifx #1\expandafter \@firstoftwo
 \else \expandafter \@secondoftwo
 \fi
}%
\providecommand \natexlab [1]{#1}%
\providecommand \enquote  [1]{``#1''}%
\providecommand \bibnamefont  [1]{#1}%
\providecommand \bibfnamefont [1]{#1}%
\providecommand \citenamefont [1]{#1}%
\providecommand \href@noop [0]{\@secondoftwo}%
\providecommand \href [0]{\begingroup \@sanitize@url \@href}%
\providecommand \@href[1]{\@@startlink{#1}\@@href}%
\providecommand \@@href[1]{\endgroup#1\@@endlink}%
\providecommand \@sanitize@url [0]{\catcode `\\12\catcode `\$12\catcode
  `\&12\catcode `\#12\catcode `\^12\catcode `\_12\catcode `\%12\relax}%
\providecommand \@@startlink[1]{}%
\providecommand \@@endlink[0]{}%
\providecommand \url  [0]{\begingroup\@sanitize@url \@url }%
\providecommand \@url [1]{\endgroup\@href {#1}{\urlprefix }}%
\providecommand \urlprefix  [0]{URL }%
\providecommand \Eprint [0]{\href }%
\providecommand \doibase [0]{https://doi.org/}%
\providecommand \selectlanguage [0]{\@gobble}%
\providecommand \bibinfo  [0]{\@secondoftwo}%
\providecommand \bibfield  [0]{\@secondoftwo}%
\providecommand \translation [1]{[#1]}%
\providecommand \BibitemOpen [0]{}%
\providecommand \bibitemStop [0]{}%
\providecommand \bibitemNoStop [0]{.\EOS\space}%
\providecommand \EOS [0]{\spacefactor3000\relax}%
\providecommand \BibitemShut  [1]{\csname bibitem#1\endcsname}%
\let\auto@bib@innerbib\@empty
\bibitem [{\citenamefont {Sumpter}(2010)}]{sumpter2010collective}%
  \BibitemOpen
  \bibfield  {author} {\bibinfo {author} {\bibfnamefont {D.~J.}\ \bibnamefont
  {Sumpter}},\ }\bibfield  {title} {\bibinfo {title} {Collective animal
  behavior},\ }in\ \href@noop {} {\emph {\bibinfo {booktitle} {Collective
  Animal Behavior}}}\ (\bibinfo  {publisher} {Princeton University Press},\
  \bibinfo {year} {2010})\BibitemShut {NoStop}%
\bibitem [{\citenamefont {Krause}\ \emph {et~al.}(2002)\citenamefont {Krause},
  \citenamefont {Krause}, \citenamefont {Ruxton},\ and\ \citenamefont
  {Ruxton}}]{krause2002living}%
  \BibitemOpen
  \bibfield  {author} {\bibinfo {author} {\bibfnamefont {J.}~\bibnamefont
  {Krause}}, \bibinfo {author} {\bibfnamefont {P.}~\bibnamefont {Krause}},
  \bibinfo {author} {\bibfnamefont {G.}~\bibnamefont {Ruxton}},\ and\ \bibinfo
  {author} {\bibfnamefont {G.}~\bibnamefont {Ruxton}},\ }\href@noop {} {\emph
  {\bibinfo {title} {Living in Groups}}},\ Oxford Series in Ecology and
  Evolution\ (\bibinfo  {publisher} {OUP Oxford},\ \bibinfo {year}
  {2002})\BibitemShut {NoStop}%
\bibitem [{\citenamefont {Vicsek}\ \emph {et~al.}(1995)\citenamefont {Vicsek},
  \citenamefont {Czir{\'o}k}, \citenamefont {Ben-Jacob}, \citenamefont
  {Cohen},\ and\ \citenamefont {Shochet}}]{vicsek1995novel}%
  \BibitemOpen
  \bibfield  {author} {\bibinfo {author} {\bibfnamefont {T.}~\bibnamefont
  {Vicsek}}, \bibinfo {author} {\bibfnamefont {A.}~\bibnamefont {Czir{\'o}k}},
  \bibinfo {author} {\bibfnamefont {E.}~\bibnamefont {Ben-Jacob}}, \bibinfo
  {author} {\bibfnamefont {I.}~\bibnamefont {Cohen}},\ and\ \bibinfo {author}
  {\bibfnamefont {O.}~\bibnamefont {Shochet}},\ }\bibfield  {title} {\bibinfo
  {title} {Novel type of phase transition in a system of self-driven
  particles},\ }\href@noop {} {\bibfield  {journal} {\bibinfo  {journal}
  {Physical Review Letters}\ }\textbf {\bibinfo {volume} {75}},\ \bibinfo
  {pages} {1226} (\bibinfo {year} {1995})}\BibitemShut {NoStop}%
\bibitem [{\citenamefont {Chung}\ \emph {et~al.}(2018)\citenamefont {Chung},
  \citenamefont {Paranjape}, \citenamefont {Dames}, \citenamefont {Shen},\ and\
  \citenamefont {Kumar}}]{chung2018survey}%
  \BibitemOpen
  \bibfield  {author} {\bibinfo {author} {\bibfnamefont {S.-J.}\ \bibnamefont
  {Chung}}, \bibinfo {author} {\bibfnamefont {A.~A.}\ \bibnamefont
  {Paranjape}}, \bibinfo {author} {\bibfnamefont {P.}~\bibnamefont {Dames}},
  \bibinfo {author} {\bibfnamefont {S.}~\bibnamefont {Shen}},\ and\ \bibinfo
  {author} {\bibfnamefont {V.}~\bibnamefont {Kumar}},\ }\bibfield  {title}
  {\bibinfo {title} {A survey on aerial swarm robotics},\ }\href@noop {}
  {\bibfield  {journal} {\bibinfo  {journal} {IEEE Transactions on Robotics}\
  }\textbf {\bibinfo {volume} {34}},\ \bibinfo {pages} {837} (\bibinfo {year}
  {2018})}\BibitemShut {NoStop}%
\bibitem [{\citenamefont {Couzin}\ \emph {et~al.}(2002)\citenamefont {Couzin},
  \citenamefont {Krause}, \citenamefont {James}, \citenamefont {Ruxton},\ and\
  \citenamefont {Franks}}]{couzin2002collective}%
  \BibitemOpen
  \bibfield  {author} {\bibinfo {author} {\bibfnamefont {I.~D.}\ \bibnamefont
  {Couzin}}, \bibinfo {author} {\bibfnamefont {J.}~\bibnamefont {Krause}},
  \bibinfo {author} {\bibfnamefont {R.}~\bibnamefont {James}}, \bibinfo
  {author} {\bibfnamefont {G.~D.}\ \bibnamefont {Ruxton}},\ and\ \bibinfo
  {author} {\bibfnamefont {N.~R.}\ \bibnamefont {Franks}},\ }\bibfield  {title}
  {\bibinfo {title} {Collective memory and spatial sorting in animal groups},\
  }\href@noop {} {\bibfield  {journal} {\bibinfo  {journal} {Journal of
  Theoretical Biology}\ }\textbf {\bibinfo {volume} {218}},\ \bibinfo {pages}
  {1} (\bibinfo {year} {2002})}\BibitemShut {NoStop}%
\bibitem [{\citenamefont {Jia}\ and\ \citenamefont
  {Vicsek}(2019)}]{jia2019modelling}%
  \BibitemOpen
  \bibfield  {author} {\bibinfo {author} {\bibfnamefont {Y.}~\bibnamefont
  {Jia}}\ and\ \bibinfo {author} {\bibfnamefont {T.}~\bibnamefont {Vicsek}},\
  }\bibfield  {title} {\bibinfo {title} {Modelling hierarchical flocking},\
  }\href@noop {} {\bibfield  {journal} {\bibinfo  {journal} {New Journal of
  Physics}\ }\textbf {\bibinfo {volume} {21}},\ \bibinfo {pages} {093048}
  (\bibinfo {year} {2019})}\BibitemShut {NoStop}%
\bibitem [{\citenamefont {Pearce}\ \emph {et~al.}(2014)\citenamefont {Pearce},
  \citenamefont {Miller}, \citenamefont {Rowlands},\ and\ \citenamefont
  {Turner}}]{pearce2014role}%
  \BibitemOpen
  \bibfield  {author} {\bibinfo {author} {\bibfnamefont {D.~J.}\ \bibnamefont
  {Pearce}}, \bibinfo {author} {\bibfnamefont {A.~M.}\ \bibnamefont {Miller}},
  \bibinfo {author} {\bibfnamefont {G.}~\bibnamefont {Rowlands}},\ and\
  \bibinfo {author} {\bibfnamefont {M.~S.}\ \bibnamefont {Turner}},\ }\bibfield
   {title} {\bibinfo {title} {Role of projection in the control of bird
  flocks},\ }\href@noop {} {\bibfield  {journal} {\bibinfo  {journal}
  {Proceedings of the National Academy of Sciences}\ }\textbf {\bibinfo
  {volume} {111}},\ \bibinfo {pages} {10422} (\bibinfo {year}
  {2014})}\BibitemShut {NoStop}%
\bibitem [{\citenamefont {Barberis}\ and\ \citenamefont
  {Peruani}(2016)}]{barberis2016large}%
  \BibitemOpen
  \bibfield  {author} {\bibinfo {author} {\bibfnamefont {L.}~\bibnamefont
  {Barberis}}\ and\ \bibinfo {author} {\bibfnamefont {F.}~\bibnamefont
  {Peruani}},\ }\bibfield  {title} {\bibinfo {title} {Large-scale patterns in a
  minimal cognitive flocking model: incidental leaders, nematic patterns, and
  aggregates},\ }\href@noop {} {\bibfield  {journal} {\bibinfo  {journal}
  {Physical Review Letters}\ }\textbf {\bibinfo {volume} {117}},\ \bibinfo
  {pages} {248001} (\bibinfo {year} {2016})}\BibitemShut {NoStop}%
\bibitem [{\citenamefont {Lavergne}\ \emph {et~al.}(2019)\citenamefont
  {Lavergne}, \citenamefont {Wendehenne}, \citenamefont {B{\"a}uerle},\ and\
  \citenamefont {Bechinger}}]{lavergne2019group}%
  \BibitemOpen
  \bibfield  {author} {\bibinfo {author} {\bibfnamefont {F.~A.}\ \bibnamefont
  {Lavergne}}, \bibinfo {author} {\bibfnamefont {H.}~\bibnamefont
  {Wendehenne}}, \bibinfo {author} {\bibfnamefont {T.}~\bibnamefont
  {B{\"a}uerle}},\ and\ \bibinfo {author} {\bibfnamefont {C.}~\bibnamefont
  {Bechinger}},\ }\bibfield  {title} {\bibinfo {title} {Group formation and
  cohesion of active particles with visual perception--dependent motility},\
  }\href@noop {} {\bibfield  {journal} {\bibinfo  {journal} {Science}\ }\textbf
  {\bibinfo {volume} {364}},\ \bibinfo {pages} {70} (\bibinfo {year}
  {2019})}\BibitemShut {NoStop}%
\bibitem [{\citenamefont {Bastien}\ and\ \citenamefont
  {Romanczuk}(2020)}]{bastien2020model}%
  \BibitemOpen
  \bibfield  {author} {\bibinfo {author} {\bibfnamefont {R.}~\bibnamefont
  {Bastien}}\ and\ \bibinfo {author} {\bibfnamefont {P.}~\bibnamefont
  {Romanczuk}},\ }\bibfield  {title} {\bibinfo {title} {A model of collective
  behavior based purely on vision},\ }\href@noop {} {\bibfield  {journal}
  {\bibinfo  {journal} {Science advances}\ }\textbf {\bibinfo {volume} {6}},\
  \bibinfo {pages} {eaay0792} (\bibinfo {year} {2020})}\BibitemShut {NoStop}%
\bibitem [{\citenamefont {Chen}\ and\ \citenamefont
  {Kolokolnikov}(2014)}]{chen2014minimal}%
  \BibitemOpen
  \bibfield  {author} {\bibinfo {author} {\bibfnamefont {Y.}~\bibnamefont
  {Chen}}\ and\ \bibinfo {author} {\bibfnamefont {T.}~\bibnamefont
  {Kolokolnikov}},\ }\bibfield  {title} {\bibinfo {title} {A minimal model of
  predator--swarm interactions},\ }\href@noop {} {\bibfield  {journal}
  {\bibinfo  {journal} {Journal of The Royal Society Interface}\ }\textbf
  {\bibinfo {volume} {11}},\ \bibinfo {pages} {20131208} (\bibinfo {year}
  {2014})}\BibitemShut {NoStop}%
\bibitem [{\citenamefont {Liu}\ \emph {et~al.}(2023)\citenamefont {Liu},
  \citenamefont {Liang}, \citenamefont {Deng},\ and\ \citenamefont
  {Zhang}}]{liu2023modeling}%
  \BibitemOpen
  \bibfield  {author} {\bibinfo {author} {\bibfnamefont {D.}~\bibnamefont
  {Liu}}, \bibinfo {author} {\bibfnamefont {Y.}~\bibnamefont {Liang}}, \bibinfo
  {author} {\bibfnamefont {J.}~\bibnamefont {Deng}},\ and\ \bibinfo {author}
  {\bibfnamefont {W.}~\bibnamefont {Zhang}},\ }\bibfield  {title} {\bibinfo
  {title} {Modeling three-dimensional bait ball collective motion},\
  }\href@noop {} {\bibfield  {journal} {\bibinfo  {journal} {Physical Review
  E}\ }\textbf {\bibinfo {volume} {107}},\ \bibinfo {pages} {014606} (\bibinfo
  {year} {2023})}\BibitemShut {NoStop}%
\bibitem [{\citenamefont {Olson}\ \emph {et~al.}(2013)\citenamefont {Olson},
  \citenamefont {Hintze}, \citenamefont {Dyer}, \citenamefont {Knoester},\ and\
  \citenamefont {Adami}}]{olson2013predator}%
  \BibitemOpen
  \bibfield  {author} {\bibinfo {author} {\bibfnamefont {R.~S.}\ \bibnamefont
  {Olson}}, \bibinfo {author} {\bibfnamefont {A.}~\bibnamefont {Hintze}},
  \bibinfo {author} {\bibfnamefont {F.~C.}\ \bibnamefont {Dyer}}, \bibinfo
  {author} {\bibfnamefont {D.~B.}\ \bibnamefont {Knoester}},\ and\ \bibinfo
  {author} {\bibfnamefont {C.}~\bibnamefont {Adami}},\ }\bibfield  {title}
  {\bibinfo {title} {Predator confusion is sufficient to evolve swarming
  behaviour},\ }\href@noop {} {\bibfield  {journal} {\bibinfo  {journal}
  {Journal of The Royal Society Interface}\ }\textbf {\bibinfo {volume} {10}},\
  \bibinfo {pages} {20130305} (\bibinfo {year} {2013})}\BibitemShut {NoStop}%
\bibitem [{\citenamefont {Sunehag}\ \emph {et~al.}(2019)\citenamefont
  {Sunehag}, \citenamefont {Lever}, \citenamefont {Liu}, \citenamefont {Merel},
  \citenamefont {Heess}, \citenamefont {Leibo}, \citenamefont {Hughes},
  \citenamefont {Eccles},\ and\ \citenamefont
  {Graepel}}]{sunehag2019reinforcement}%
  \BibitemOpen
  \bibfield  {author} {\bibinfo {author} {\bibfnamefont {P.}~\bibnamefont
  {Sunehag}}, \bibinfo {author} {\bibfnamefont {G.}~\bibnamefont {Lever}},
  \bibinfo {author} {\bibfnamefont {S.}~\bibnamefont {Liu}}, \bibinfo {author}
  {\bibfnamefont {J.}~\bibnamefont {Merel}}, \bibinfo {author} {\bibfnamefont
  {N.}~\bibnamefont {Heess}}, \bibinfo {author} {\bibfnamefont {J.~Z.}\
  \bibnamefont {Leibo}}, \bibinfo {author} {\bibfnamefont {E.}~\bibnamefont
  {Hughes}}, \bibinfo {author} {\bibfnamefont {T.}~\bibnamefont {Eccles}},\
  and\ \bibinfo {author} {\bibfnamefont {T.}~\bibnamefont {Graepel}},\
  }\bibfield  {title} {\bibinfo {title} {Reinforcement learning agents acquire
  flocking and symbiotic behaviour in simulated ecosystems},\ }in\ \href@noop
  {} {\emph {\bibinfo {booktitle} {ALIFE 2019: The 2019 Conference on
  Artificial Life}}}\ (\bibinfo {organization} {MIT Press},\ \bibinfo {year}
  {2019})\ pp.\ \bibinfo {pages} {103--110}\BibitemShut {NoStop}%
\bibitem [{\citenamefont {Hahn}\ \emph {et~al.}(2019)\citenamefont {Hahn},
  \citenamefont {Phan}, \citenamefont {Gabor}, \citenamefont {Belzner},\ and\
  \citenamefont {Linnhoff-Popien}}]{hahn2019emergent}%
  \BibitemOpen
  \bibfield  {author} {\bibinfo {author} {\bibfnamefont {C.}~\bibnamefont
  {Hahn}}, \bibinfo {author} {\bibfnamefont {T.}~\bibnamefont {Phan}}, \bibinfo
  {author} {\bibfnamefont {T.}~\bibnamefont {Gabor}}, \bibinfo {author}
  {\bibfnamefont {L.}~\bibnamefont {Belzner}},\ and\ \bibinfo {author}
  {\bibfnamefont {C.}~\bibnamefont {Linnhoff-Popien}},\ }\bibfield  {title}
  {\bibinfo {title} {Emergent escape-based flocking behavior using multi-agent
  reinforcement learning},\ }in\ \href@noop {} {\emph {\bibinfo {booktitle}
  {ALIFE 2019: The 2019 Conference on Artificial Life}}}\ (\bibinfo
  {organization} {MIT Press},\ \bibinfo {year} {2019})\ pp.\ \bibinfo {pages}
  {598--605}\BibitemShut {NoStop}%
\bibitem [{\citenamefont {Durve}\ \emph {et~al.}(2020)\citenamefont {Durve},
  \citenamefont {Peruani},\ and\ \citenamefont {Celani}}]{durve2020learning}%
  \BibitemOpen
  \bibfield  {author} {\bibinfo {author} {\bibfnamefont {M.}~\bibnamefont
  {Durve}}, \bibinfo {author} {\bibfnamefont {F.}~\bibnamefont {Peruani}},\
  and\ \bibinfo {author} {\bibfnamefont {A.}~\bibnamefont {Celani}},\
  }\bibfield  {title} {\bibinfo {title} {Learning to flock through
  reinforcement},\ }\href@noop {} {\bibfield  {journal} {\bibinfo  {journal}
  {Physical Review E}\ }\textbf {\bibinfo {volume} {102}},\ \bibinfo {pages}
  {012601} (\bibinfo {year} {2020})}\BibitemShut {NoStop}%
\bibitem [{\citenamefont {Monter}\ \emph {et~al.}(2023)\citenamefont {Monter},
  \citenamefont {Heuthe}, \citenamefont {Panizon},\ and\ \citenamefont
  {Bechinger}}]{monter2023dynamics}%
  \BibitemOpen
  \bibfield  {author} {\bibinfo {author} {\bibfnamefont {S.}~\bibnamefont
  {Monter}}, \bibinfo {author} {\bibfnamefont {V.-L.}\ \bibnamefont {Heuthe}},
  \bibinfo {author} {\bibfnamefont {E.}~\bibnamefont {Panizon}},\ and\ \bibinfo
  {author} {\bibfnamefont {C.}~\bibnamefont {Bechinger}},\ }\bibfield  {title}
  {\bibinfo {title} {Dynamics and risk sharing in groups of selfish
  individuals},\ }\href@noop {} {\bibfield  {journal} {\bibinfo  {journal}
  {Journal of Theoretical Biology}\ }\textbf {\bibinfo {volume} {562}},\
  \bibinfo {pages} {111433} (\bibinfo {year} {2023})}\BibitemShut {NoStop}%
\bibitem [{\citenamefont {Sutton}\ and\ \citenamefont
  {Barto}(2018)}]{sutton2018reinforcement}%
  \BibitemOpen
  \bibfield  {author} {\bibinfo {author} {\bibfnamefont {R.~S.}\ \bibnamefont
  {Sutton}}\ and\ \bibinfo {author} {\bibfnamefont {A.~G.}\ \bibnamefont
  {Barto}},\ }\href@noop {} {\emph {\bibinfo {title} {Reinforcement learning:
  An introduction}}}\ (\bibinfo  {publisher} {MIT press},\ \bibinfo {year}
  {2018})\BibitemShut {NoStop}%
\bibitem [{\citenamefont {Mui{\~n}os-Landin}\ \emph {et~al.}(2021)\citenamefont
  {Mui{\~n}os-Landin}, \citenamefont {Fischer}, \citenamefont {Holubec},\ and\
  \citenamefont {Cichos}}]{muinos2021reinforcement}%
  \BibitemOpen
  \bibfield  {author} {\bibinfo {author} {\bibfnamefont {S.}~\bibnamefont
  {Mui{\~n}os-Landin}}, \bibinfo {author} {\bibfnamefont {A.}~\bibnamefont
  {Fischer}}, \bibinfo {author} {\bibfnamefont {V.}~\bibnamefont {Holubec}},\
  and\ \bibinfo {author} {\bibfnamefont {F.}~\bibnamefont {Cichos}},\
  }\bibfield  {title} {\bibinfo {title} {Reinforcement learning with artificial
  microswimmers},\ }\href@noop {} {\bibfield  {journal} {\bibinfo  {journal}
  {Science Robotics}\ }\textbf {\bibinfo {volume} {6}},\ \bibinfo {pages}
  {eabd9285} (\bibinfo {year} {2021})}\BibitemShut {NoStop}%
\bibitem [{\citenamefont {Nasiri}\ and\ \citenamefont
  {Liebchen}(2022)}]{nasiri2022reinforcement}%
  \BibitemOpen
  \bibfield  {author} {\bibinfo {author} {\bibfnamefont {M.}~\bibnamefont
  {Nasiri}}\ and\ \bibinfo {author} {\bibfnamefont {B.}~\bibnamefont
  {Liebchen}},\ }\bibfield  {title} {\bibinfo {title} {Reinforcement learning
  of optimal active particle navigation},\ }\href@noop {} {\bibfield  {journal}
  {\bibinfo  {journal} {New Journal of Physics}\ }\textbf {\bibinfo {volume}
  {24}},\ \bibinfo {pages} {073042} (\bibinfo {year} {2022})}\BibitemShut
  {NoStop}%
\bibitem [{\citenamefont {Kaiser}\ and\ \citenamefont
  {Hamann}(2022)}]{kaiser2022innate}%
  \BibitemOpen
  \bibfield  {author} {\bibinfo {author} {\bibfnamefont {T.~K.}\ \bibnamefont
  {Kaiser}}\ and\ \bibinfo {author} {\bibfnamefont {H.}~\bibnamefont
  {Hamann}},\ }\bibfield  {title} {\bibinfo {title} {Innate motivation for
  robot swarms by minimizing surprise: From simple simulations to real-world
  experiments},\ }\href@noop {} {\bibfield  {journal} {\bibinfo  {journal}
  {IEEE Transactions on Robotics}\ }\textbf {\bibinfo {volume} {38}},\ \bibinfo
  {pages} {3582} (\bibinfo {year} {2022})}\BibitemShut {NoStop}%
\bibitem [{\citenamefont {Nelson}\ \emph {et~al.}(2009)\citenamefont {Nelson},
  \citenamefont {Barlow},\ and\ \citenamefont {Doitsidis}}]{nelson2009fitness}%
  \BibitemOpen
  \bibfield  {author} {\bibinfo {author} {\bibfnamefont {A.~L.}\ \bibnamefont
  {Nelson}}, \bibinfo {author} {\bibfnamefont {G.~J.}\ \bibnamefont {Barlow}},\
  and\ \bibinfo {author} {\bibfnamefont {L.}~\bibnamefont {Doitsidis}},\
  }\bibfield  {title} {\bibinfo {title} {Fitness functions in evolutionary
  robotics: A survey and analysis},\ }\href@noop {} {\bibfield  {journal}
  {\bibinfo  {journal} {Robotics and Autonomous Systems}\ }\textbf {\bibinfo
  {volume} {57}},\ \bibinfo {pages} {345} (\bibinfo {year} {2009})}\BibitemShut
  {NoStop}%
\bibitem [{\citenamefont {Hamilton}(1971)}]{hamilton1971geometry}%
  \BibitemOpen
  \bibfield  {author} {\bibinfo {author} {\bibfnamefont {W.~D.}\ \bibnamefont
  {Hamilton}},\ }\bibfield  {title} {\bibinfo {title} {Geometry for the selfish
  herd},\ }\href@noop {} {\bibfield  {journal} {\bibinfo  {journal} {Journal of
  Theoretical Biology}\ }\textbf {\bibinfo {volume} {31}},\ \bibinfo {pages}
  {295} (\bibinfo {year} {1971})}\BibitemShut {NoStop}%
\bibitem [{\citenamefont {Mordatch}\ and\ \citenamefont
  {Abbeel}(2017)}]{mordatch2017emergence}%
  \BibitemOpen
  \bibfield  {author} {\bibinfo {author} {\bibfnamefont {I.}~\bibnamefont
  {Mordatch}}\ and\ \bibinfo {author} {\bibfnamefont {P.}~\bibnamefont
  {Abbeel}},\ }\bibfield  {title} {\bibinfo {title} {Emergence of grounded
  compositional language in multi-agent populations},\ }\href@noop {}
  {\bibfield  {journal} {\bibinfo  {journal} {arXiv preprint arXiv:1703.04908}\
  } (\bibinfo {year} {2017})}\BibitemShut {NoStop}%
\bibitem [{\citenamefont {Lowe}\ \emph {et~al.}(2017)\citenamefont {Lowe},
  \citenamefont {Wu}, \citenamefont {Tamar}, \citenamefont {Harb},
  \citenamefont {Abbeel},\ and\ \citenamefont {Mordatch}}]{lowe2017multi}%
  \BibitemOpen
  \bibfield  {author} {\bibinfo {author} {\bibfnamefont {R.}~\bibnamefont
  {Lowe}}, \bibinfo {author} {\bibfnamefont {Y.}~\bibnamefont {Wu}}, \bibinfo
  {author} {\bibfnamefont {A.}~\bibnamefont {Tamar}}, \bibinfo {author}
  {\bibfnamefont {J.}~\bibnamefont {Harb}}, \bibinfo {author} {\bibfnamefont
  {P.}~\bibnamefont {Abbeel}},\ and\ \bibinfo {author} {\bibfnamefont
  {I.}~\bibnamefont {Mordatch}},\ }\bibfield  {title} {\bibinfo {title}
  {Multi-agent actor-critic for mixed cooperative-competitive environments},\
  }\href@noop {} {\bibfield  {journal} {\bibinfo  {journal} {Neural Information
  Processing Systems (NIPS)}\ } (\bibinfo {year} {2017})}\BibitemShut {NoStop}%
\bibitem [{\citenamefont {Frenkel}\ and\ \citenamefont
  {Smit}(2001)}]{frenkel2001understanding}%
  \BibitemOpen
  \bibfield  {author} {\bibinfo {author} {\bibfnamefont {D.}~\bibnamefont
  {Frenkel}}\ and\ \bibinfo {author} {\bibfnamefont {B.}~\bibnamefont {Smit}},\
  }\href@noop {} {\emph {\bibinfo {title} {Understanding molecular simulation:
  from algorithms to applications}}},\ Vol.~\bibinfo {volume} {1}\ (\bibinfo
  {publisher} {Elsevier},\ \bibinfo {year} {2001})\BibitemShut {NoStop}%
\bibitem [{\citenamefont {Ballerini}\ \emph {et~al.}(2008)\citenamefont
  {Ballerini}, \citenamefont {Cabibbo}, \citenamefont {Candelier},
  \citenamefont {Cavagna}, \citenamefont {Cisbani}, \citenamefont {Giardina},
  \citenamefont {Lecomte}, \citenamefont {Orlandi}, \citenamefont {Parisi},
  \citenamefont {Procaccini} \emph {et~al.}}]{ballerini2008interaction}%
  \BibitemOpen
  \bibfield  {author} {\bibinfo {author} {\bibfnamefont {M.}~\bibnamefont
  {Ballerini}}, \bibinfo {author} {\bibfnamefont {N.}~\bibnamefont {Cabibbo}},
  \bibinfo {author} {\bibfnamefont {R.}~\bibnamefont {Candelier}}, \bibinfo
  {author} {\bibfnamefont {A.}~\bibnamefont {Cavagna}}, \bibinfo {author}
  {\bibfnamefont {E.}~\bibnamefont {Cisbani}}, \bibinfo {author} {\bibfnamefont
  {I.}~\bibnamefont {Giardina}}, \bibinfo {author} {\bibfnamefont
  {V.}~\bibnamefont {Lecomte}}, \bibinfo {author} {\bibfnamefont
  {A.}~\bibnamefont {Orlandi}}, \bibinfo {author} {\bibfnamefont
  {G.}~\bibnamefont {Parisi}}, \bibinfo {author} {\bibfnamefont
  {A.}~\bibnamefont {Procaccini}}, \emph {et~al.},\ }\bibfield  {title}
  {\bibinfo {title} {Interaction ruling animal collective behavior depends on
  topological rather than metric distance: Evidence from a field study},\
  }\href@noop {} {\bibfield  {journal} {\bibinfo  {journal} {Proceedings of the
  National Academy of Sciences}\ }\textbf {\bibinfo {volume} {105}},\ \bibinfo
  {pages} {1232} (\bibinfo {year} {2008})}\BibitemShut {NoStop}%
\bibitem [{\citenamefont {Moussa{\"\i}d}\ \emph {et~al.}(2011)\citenamefont
  {Moussa{\"\i}d}, \citenamefont {Helbing},\ and\ \citenamefont
  {Theraulaz}}]{moussaid2011simple}%
  \BibitemOpen
  \bibfield  {author} {\bibinfo {author} {\bibfnamefont {M.}~\bibnamefont
  {Moussa{\"\i}d}}, \bibinfo {author} {\bibfnamefont {D.}~\bibnamefont
  {Helbing}},\ and\ \bibinfo {author} {\bibfnamefont {G.}~\bibnamefont
  {Theraulaz}},\ }\bibfield  {title} {\bibinfo {title} {How simple rules
  determine pedestrian behavior and crowd disasters},\ }\href@noop {}
  {\bibfield  {journal} {\bibinfo  {journal} {Proceedings of the National
  Academy of Sciences}\ }\textbf {\bibinfo {volume} {108}},\ \bibinfo {pages}
  {6884} (\bibinfo {year} {2011})}\BibitemShut {NoStop}%
\bibitem [{\citenamefont {H{\"u}ttenrauch}\ \emph {et~al.}(2019)\citenamefont
  {H{\"u}ttenrauch}, \citenamefont {Adrian}, \citenamefont {Neumann} \emph
  {et~al.}}]{huttenrauch2019deep}%
  \BibitemOpen
  \bibfield  {author} {\bibinfo {author} {\bibfnamefont {M.}~\bibnamefont
  {H{\"u}ttenrauch}}, \bibinfo {author} {\bibfnamefont {S.}~\bibnamefont
  {Adrian}}, \bibinfo {author} {\bibfnamefont {G.}~\bibnamefont {Neumann}},
  \emph {et~al.},\ }\bibfield  {title} {\bibinfo {title} {Deep reinforcement
  learning for swarm systems},\ }\href@noop {} {\bibfield  {journal} {\bibinfo
  {journal} {Journal of Machine Learning Research}\ }\textbf {\bibinfo {volume}
  {20}},\ \bibinfo {pages} {1} (\bibinfo {year} {2019})}\BibitemShut {NoStop}%
\bibitem [{\citenamefont {Sutton}\ \emph {et~al.}(1999)\citenamefont {Sutton},
  \citenamefont {McAllester}, \citenamefont {Singh},\ and\ \citenamefont
  {Mansour}}]{sutton1999policy}%
  \BibitemOpen
  \bibfield  {author} {\bibinfo {author} {\bibfnamefont {R.~S.}\ \bibnamefont
  {Sutton}}, \bibinfo {author} {\bibfnamefont {D.}~\bibnamefont {McAllester}},
  \bibinfo {author} {\bibfnamefont {S.}~\bibnamefont {Singh}},\ and\ \bibinfo
  {author} {\bibfnamefont {Y.}~\bibnamefont {Mansour}},\ }\bibfield  {title}
  {\bibinfo {title} {Policy gradient methods for reinforcement learning with
  function approximation},\ }\href@noop {} {\bibfield  {journal} {\bibinfo
  {journal} {Advances in Neural Information Processing Systems}\ }\textbf
  {\bibinfo {volume} {12}} (\bibinfo {year} {1999})}\BibitemShut {NoStop}%
\bibitem [{\citenamefont {Reynolds}(1987)}]{reynolds1987flocks}%
  \BibitemOpen
  \bibfield  {author} {\bibinfo {author} {\bibfnamefont {C.~W.}\ \bibnamefont
  {Reynolds}},\ }\bibfield  {title} {\bibinfo {title} {Flocks, herds and
  schools: A distributed behavioral model},\ }in\ \href@noop {} {\emph
  {\bibinfo {booktitle} {Proceedings of the 14th annual conference on Computer
  graphics and interactive techniques}}}\ (\bibinfo {year} {1987})\ pp.\
  \bibinfo {pages} {25--34}\BibitemShut {NoStop}%
\bibitem [{\citenamefont {Huffaker}\ \emph {et~al.}(1958)\citenamefont
  {Huffaker} \emph {et~al.}}]{huffaker1958experimental}%
  \BibitemOpen
  \bibfield  {author} {\bibinfo {author} {\bibfnamefont {C.}~\bibnamefont
  {Huffaker}} \emph {et~al.},\ }\bibfield  {title} {\bibinfo {title}
  {Experimental studies on predation: dispersion factors and predator-prey
  oscillations},\ }\href@noop {} {\bibfield  {journal} {\bibinfo  {journal}
  {Hilgardia}\ }\textbf {\bibinfo {volume} {27}},\ \bibinfo {pages} {343}
  (\bibinfo {year} {1958})}\BibitemShut {NoStop}%
\bibitem [{\citenamefont {Major}(1978)}]{major1978predator}%
  \BibitemOpen
  \bibfield  {author} {\bibinfo {author} {\bibfnamefont {P.~F.}\ \bibnamefont
  {Major}},\ }\bibfield  {title} {\bibinfo {title} {Predator-prey interactions
  in two schooling fishes, caranx ignobilis and stolephorus purpureus},\
  }\href@noop {} {\bibfield  {journal} {\bibinfo  {journal} {Animal Behaviour}\
  }\textbf {\bibinfo {volume} {26}},\ \bibinfo {pages} {760} (\bibinfo {year}
  {1978})}\BibitemShut {NoStop}%
\bibitem [{\citenamefont {Duffield}\ and\ \citenamefont
  {Ioannou}(2017)}]{duffield2017marginal}%
  \BibitemOpen
  \bibfield  {author} {\bibinfo {author} {\bibfnamefont {C.}~\bibnamefont
  {Duffield}}\ and\ \bibinfo {author} {\bibfnamefont {C.~C.}\ \bibnamefont
  {Ioannou}},\ }\bibfield  {title} {\bibinfo {title} {Marginal predation: do
  encounter or confusion effects explain the targeting of prey group edges?},\
  }\href@noop {} {\bibfield  {journal} {\bibinfo  {journal} {Behavioral
  Ecology}\ }\textbf {\bibinfo {volume} {28}},\ \bibinfo {pages} {1283}
  (\bibinfo {year} {2017})}\BibitemShut {NoStop}%
\bibitem [{\citenamefont {Romenskyy}\ \emph {et~al.}(2020)\citenamefont
  {Romenskyy}, \citenamefont {Herbert-Read}, \citenamefont {Ioannou},
  \citenamefont {Szorkovszky}, \citenamefont {Ward},\ and\ \citenamefont
  {Sumpter}}]{romenskyy2020quantifying}%
  \BibitemOpen
  \bibfield  {author} {\bibinfo {author} {\bibfnamefont {M.}~\bibnamefont
  {Romenskyy}}, \bibinfo {author} {\bibfnamefont {J.~E.}\ \bibnamefont
  {Herbert-Read}}, \bibinfo {author} {\bibfnamefont {C.~C.}\ \bibnamefont
  {Ioannou}}, \bibinfo {author} {\bibfnamefont {A.}~\bibnamefont
  {Szorkovszky}}, \bibinfo {author} {\bibfnamefont {A.~J.}\ \bibnamefont
  {Ward}},\ and\ \bibinfo {author} {\bibfnamefont {D.~J.}\ \bibnamefont
  {Sumpter}},\ }\bibfield  {title} {\bibinfo {title} {Quantifying the structure
  and dynamics of fish shoals under predation threat in three dimensions},\
  }\href@noop {} {\bibfield  {journal} {\bibinfo  {journal} {Behavioral
  Ecology}\ }\textbf {\bibinfo {volume} {31}},\ \bibinfo {pages} {311}
  (\bibinfo {year} {2020})}\BibitemShut {NoStop}%
\bibitem [{\citenamefont {Krakauer}(1995)}]{krakauer1995groups}%
  \BibitemOpen
  \bibfield  {author} {\bibinfo {author} {\bibfnamefont {D.~C.}\ \bibnamefont
  {Krakauer}},\ }\bibfield  {title} {\bibinfo {title} {Groups confuse predators
  by exploiting perceptual bottlenecks: a connectionist model of the confusion
  effect},\ }\href@noop {} {\bibfield  {journal} {\bibinfo  {journal}
  {Behavioral Ecology and Sociobiology}\ }\textbf {\bibinfo {volume} {36}},\
  \bibinfo {pages} {421} (\bibinfo {year} {1995})}\BibitemShut {NoStop}%
\bibitem [{\citenamefont {Parrish}\ and\ \citenamefont
  {Edelstein-Keshet}(1999)}]{parrish1999complexity}%
  \BibitemOpen
  \bibfield  {author} {\bibinfo {author} {\bibfnamefont {J.~K.}\ \bibnamefont
  {Parrish}}\ and\ \bibinfo {author} {\bibfnamefont {L.}~\bibnamefont
  {Edelstein-Keshet}},\ }\bibfield  {title} {\bibinfo {title} {Complexity,
  pattern, and evolutionary trade-offs in animal aggregation},\ }\href@noop {}
  {\bibfield  {journal} {\bibinfo  {journal} {Science}\ }\textbf {\bibinfo
  {volume} {284}},\ \bibinfo {pages} {99} (\bibinfo {year} {1999})}\BibitemShut
  {NoStop}%
\bibitem [{\citenamefont {Franks}\ \emph {et~al.}(2016)\citenamefont {Franks},
  \citenamefont {Worley}, \citenamefont {Grant}, \citenamefont {Gorman},
  \citenamefont {Vizard}, \citenamefont {Plackett}, \citenamefont {Doran},
  \citenamefont {Gamble}, \citenamefont {Stumpe},\ and\ \citenamefont
  {Sendova-Franks}}]{franks2016social}%
  \BibitemOpen
  \bibfield  {author} {\bibinfo {author} {\bibfnamefont {N.~R.}\ \bibnamefont
  {Franks}}, \bibinfo {author} {\bibfnamefont {A.}~\bibnamefont {Worley}},
  \bibinfo {author} {\bibfnamefont {K.~A.}\ \bibnamefont {Grant}}, \bibinfo
  {author} {\bibfnamefont {A.~R.}\ \bibnamefont {Gorman}}, \bibinfo {author}
  {\bibfnamefont {V.}~\bibnamefont {Vizard}}, \bibinfo {author} {\bibfnamefont
  {H.}~\bibnamefont {Plackett}}, \bibinfo {author} {\bibfnamefont
  {C.}~\bibnamefont {Doran}}, \bibinfo {author} {\bibfnamefont {M.~L.}\
  \bibnamefont {Gamble}}, \bibinfo {author} {\bibfnamefont {M.~C.}\
  \bibnamefont {Stumpe}},\ and\ \bibinfo {author} {\bibfnamefont {A.~B.}\
  \bibnamefont {Sendova-Franks}},\ }\bibfield  {title} {\bibinfo {title}
  {Social behaviour and collective motion in plant-animal worms},\ }\href@noop
  {} {\bibfield  {journal} {\bibinfo  {journal} {Proceedings of the Royal
  Society B: Biological Sciences}\ }\textbf {\bibinfo {volume} {283}},\
  \bibinfo {pages} {20152946} (\bibinfo {year} {2016})}\BibitemShut {NoStop}%
\bibitem [{\citenamefont {Zhao}(2024)}]{zhao2024reinforcement}%
  \BibitemOpen
  \bibfield  {author} {\bibinfo {author} {\bibfnamefont {S.}~\bibnamefont
  {Zhao}},\ }\href@noop {} {\emph {\bibinfo {title} {Mathematical Foundations
  of Reinforcement Learning}}}\ (\bibinfo  {publisher} {Tsinghua University
  Press and Springer Nature Press},\ \bibinfo {year} {2024})\BibitemShut
  {NoStop}%
\bibitem [{\citenamefont {Zhang}\ \emph {et~al.}(2021)\citenamefont {Zhang},
  \citenamefont {Yang},\ and\ \citenamefont {Ba{\c{s}}ar}}]{zhang2021multi}%
  \BibitemOpen
  \bibfield  {author} {\bibinfo {author} {\bibfnamefont {K.}~\bibnamefont
  {Zhang}}, \bibinfo {author} {\bibfnamefont {Z.}~\bibnamefont {Yang}},\ and\
  \bibinfo {author} {\bibfnamefont {T.}~\bibnamefont {Ba{\c{s}}ar}},\
  }\bibfield  {title} {\bibinfo {title} {Multi-agent reinforcement learning: A
  selective overview of theories and algorithms},\ }\href@noop {} {\bibfield
  {journal} {\bibinfo  {journal} {Handbook of reinforcement learning and
  control}\ ,\ \bibinfo {pages} {321}} (\bibinfo {year} {2021})}\BibitemShut
  {NoStop}%
\end{thebibliography}%

\end{document}